\documentclass[sigconf]{acmart}

\renewcommand\footnotetextcopyrightpermission[1]{} 

\AtBeginDocument{%
  \providecommand\BibTeX{{%
    \normalfont B\kern-0.5em{\scshape i\kern-0.25em b}\kern-0.8em\TeX}}}

\usepackage{amsmath}
\newtheorem{define}{Definition}
\usepackage{bm}
\usepackage{braket} 
\usepackage[caption=false,font=large]{subfig}
\usepackage[capitalize]{cleveref}
\usepackage[shortlabels]{enumitem}    
\usepackage{multirow}

\usepackage{amsthm}
\makeatletter
\def\thmheadbrackets#1#2#3{%
  \thmname{#1}\thmnumber{\@ifnotempty{#1}{ }\@upn{#2}}%
  \thmnote{ {\the\thm@notefont[#3]}}}
\makeatother
\newtheoremstyle{brackets}
  {}
  {}
  {\itshape}
  {}
  {\bfseries}
  {.}
  { }
  {\thmheadbrackets{#1}{#2}{#3}}
\theoremstyle{brackets}
\newtheorem{interpret}{Interpretation}

\begin{document}

\title{Quantum Remote Entanglement for Medium-Free Secure Communication?}

\author{Wesley Joon-Wie Tann}
\affiliation{%
  \institution{National University of Singapore}
  \country{}   }
\email{wesleyjtann@u.nus.edu}


\begin{abstract}
Present-day quantum communication predominantly depends on trusted relays (e.g., quantum repeaters, low-Earth-orbit satellite) connected by optical fiber cables to transmit information. 
However, recent evidence supports a decades-old concept that quantum entanglement, harnessed by current quantum communication systems, does not necessarily rely on a physical relay medium. 
In modern quantum communication networks, this trusted relay infrastructure is (1) susceptible to security attacks, (2) limited by the channel capacity, (3) subject to decoherence loss, and (4) expensive to set up. The instantaneous and faster-than-light activities of quantum entanglement occurring in quantum communication have suggested guidance by some non-locality nature. 
On the contrary, neither ground nor space-relays have shown or been demonstrated to embody it. 
It is proposed in this paper that the non-locality nature of quantum theory governs quantum entanglement; elementary particles, components of a universal quantum body, can achieve remote entanglement regardless of a physical medium or spatial proximity. 
Evidence and theory supporting remote entanglement in superconducting quantum systems (entanglement fidelities for communication in particular) are presented. 
One such particle, the photon, representing a basic unit of quantum information, qubit $\ket{\psi} = \alpha \ket{0} + \beta \ket{1}$, consists of real continuous values in complex numbers $(\alpha, \beta)$ with infinite precision. These values $(\alpha, \beta)$ can account for the distinctiveness of qubits and result in an identity $QuID$ that possibly supports remote entanglement. 
New approaches to medium-free secure quantum communication are suggested by running simulations and actual quantum computations on a quantum circuit.


\end{abstract}

\ccsdesc[500]{Applied computing~Telecommunications}
\ccsdesc[500]{Networks~Network security}
\ccsdesc[500]{Networks~Wireless access networks}

\keywords{Remote quantum entanglement, Wireless medium-free communication, Future network security}


\maketitle

\section{Introduction}
Quantum communication leverages the fundamental properties of quantum mechanics for information transmission, sending basic units of information in quantum bits from one quantum processor to another. Most modern quantum communication systems are built upon physical networks using telecommunication optical fiber infrastructures that rely on trusted quantum repeaters to relay information.
Even though quantum communication protocols are cryptographically secure~\cite{Lo_2014}, they are implemented on top of a fiber-optical physical layer---reliant on trusted relay nodes in the network---exposing the communication to security vulnerabilities, such as denial-of-service attacks~\cite{5749282}. 
With the advent of quantum communication, we ask if it is possible to devise a communication system that delivers information directly between parties, thus avoiding vulnerabilities of the trusted relay network architecture.

Existing quantum communication networks are generally satellite-based or fiber-optic-based quantum distribution systems. On the one hand, satellite communication methods~\cite{PhysRevLett.115.040502,Liao_2017} typically create a network of satellites and ground stations to perform the transmission of quantum states. In the first space-to-ground quantum communication link, a satellite sent a pair of entangled photons in free space to ground stations up to 1200 kilometers (km) apart for quantum key distribution. One year later, the same satellite, acting as a trusted relay, distributed a secure key between two intercontinental ground stations separated by 7600 km on Earth~\cite{PhysRevLett.120.030501}.
On the other hand, optical networks~\cite{duan2001long,elliott2002building,gisin2007quantum,Chen:2010} transport quantum information from one end node to another node in the network by propagating photons through optical fibers. Standard telecommunication fibers can be used for such purposes. 
These networks require trusted quantum repeaters to establish communication over extended distances. 

However, entangled quantum states degrade as they pass through air or optical fiber, suffering the loss of entanglement. 
Moreover, both existing types of quantum communication networks rely on physical trusted relays, presenting a common set of challenges. $(1)$ Quantum communication networks built on any physical infrastructure are vulnerable to security threats such as denial-of-service attacks.
$(2)$ The bandwidth of any physical channel (e.g., fiber optic cables) poses an inherent limit on the maximum data transmission rate.
$(3)$ The atmospheric effects on quantum transmission loss and decoherence, diminishing long-distance communication quality, pose yet another challenge for physical quantum communication systems.
$(4)$ Cost constraints are a real consideration in the setup of a physical quantum network infrastructure, and communication systems have to be economically reasonable to scale quantum communication efficiently.
In this paper, we propose a departing view from the latest understanding of quantum interactions, which, by extension, overcomes the challenges of communication that accompany physical transmission networks.
Every quantum particle is described by a dense collection of real numbers arbitrarily close in value to an infinite number of neighbors. One such number is the wavelength $\lambda = \hbar / p$ of infinite precision~\cite{de1924recherches}, relating to the Planck's constant $\hbar$ and its momentum $p$. It implies that particles are identifiable given resolving power.
Our interpretation, the \textit{Quantum Remote Entanglement} (QRE), theorizes that this distinct wavelength can be viewed as an identifier $QuID$ and is fundamental in the interactions among particles. 
Based on this view, harmonizing with the inherently non-local nature of quantum theory, could it be possible for any two specific particles to establish remote entanglement and transmit quantum information between them, bypassing the need for a physical transmission medium?


The remote entanglement of two separate quantum systems has been repeatedly realized in the form of mediated entanglement~\cite{campagne2018deterministic,humphreys2018deterministic,kurpiers2018deterministic}. Non-entangled, spatially apart qubits are unified by applying entangling measurements that project them into a maximum superposition entangled state. In some cases, using photons~\cite{hensen2015loophole,zhong2019violating}, the coupling of distant quantum systems requires the quantum system to be coupled to the photons to perform remote entanglement. For other instances in which the quantum systems are not coupled to the qubits, a universal medium is needed for coupling the remote quantum systems. One such method~\cite{bienfait2019phonon} proposed the emission and capture of itinerant surface acoustic wave phonons, enabling the quantum entanglement of two superconducting qubits. These current mediated methods achieve remote entanglement of separate non-entangled qubits via an intermediary.

\begin{figure}[htbp]
        \centering
        \includegraphics[width=.6\linewidth]{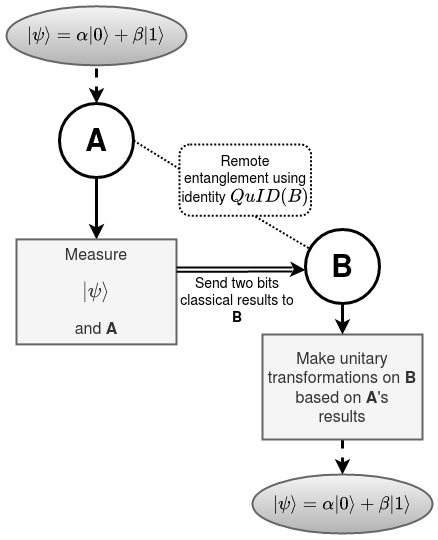}
        \caption{Quantum remote entanglement for medium-free secure communication. The dashed line represents the communication of a prepared quantum state $\ket{\psi}$, the dotted line a remote entanglement of particles Alice $(A)$ and Bob $(B)$, the single solid line some quantum information, and the double solid line a classical pair of bits.}
        \label{fig:remoteteleport}
    \end{figure}

Using photons, an elementary particle that is the quantum of the electromagnetic field, we let it represent a quantum bit $\ket{\psi} = \alpha \ket{0} + \beta \ket{1}$, where $\alpha, \beta \in \mathbb{C}$ are complex numbers with real continuous values of infinite precision.
In this case, each qubit $i$ is distinguishable by its $\alpha$ and $\beta$ values, and we can define the identity as $QuID (i) = (\alpha_i, \beta_i)$.
Now, assuming the QRE interpretation and that qubits represented by photon particles could, in fact, remotely entangle using their $QuID$s, they could interact with each other from any distance.
For example, Alice prepared a qubit in a quantum state $\ket{\psi}$ to communicate to Bob. In addition, while Bob has his qubit $B$, Alice has another qubit $A$ and the identity $QuID(B)$ of Bob's qubit. Quantum tomography~\cite{lvovsky2009continuous} can be used to determine the identity of Bob's qubit reliably. 
Thus, she can remotely entangle her qubit $A$ with Bob's qubit $B$ using $QuID(B)$. Once $A$ and $B$ are entangled into an Einstein-Podolsky-Rosen (EPR)~\cite{einstein1935can} pair, the canonical teleportation protocol~\cite{PhysRevLett.70.1895} can be initiated, allowing secure quantum communication solely based on QRE without any physical channel for the transmission of information (see Figure~\ref{fig:remoteteleport}).

Experiments on quantum circuits, adding evidential value, suggest the ability of our approach to realize medium-free secure quantum communication.
It provides a step in proving this concept and validating the approach. 
Our purpose is to present a proposal that has the potential for future quantum communication. 
We run both simulations and actual quantum computations on a quantum circuit, where Alice sends a qubit $\ket{\psi}$ in a prepared state to Bob, using the concept of QRE and the teleportation protocol.

In the first simulation, reproducing quantum state vectors, the circuit successfully transmitted Alice's prepared communication state $(\alpha, \beta) = (-0.57659 + 0.24170i, -0.59478 - 0.50532i)$ to Bob (see Figure~\ref{fig:statevec_2}). 
Next, we run the noisy quantum circuit. We first prepare $\ket{\psi}$ by putting it into a superposition state of $\ket{0}$. The goal is to teleport $\ket{\psi}$ to Bob and have him measure the teleported $\ket{\psi}$ to get either a binary bit $\ket{0}$ or $\ket{1}$. If the binary bit measured by Bob turns out to be $\ket{0}$, it indicates that $\ket{\psi}$ has been correctly transmitted. Since the circuit is noisy, we sample it $1024$ times to get a frequency distribution. 
We repeat the same simulation for $\ket{\psi}$ in state $\ket{1}$ (see Figures \ref{fig:simulateket0} and \ref{fig:simulateket1}). In these simulations, we see both $\ket{0}$ and $\ket{1}$ measured 100\% of the time as the simulation is error-free.

Finally, we run the noisy circuit on a real quantum computer, with errors in the computations and qubits due to environmental impacts. We also sample it $1024$ times to get a frequency distribution (see Figure~\ref{fig:realcompute}). As expected, there are errors in the measurements, returning experimental error rates of $8.9\%$ and $5.6\%$ for $\ket{0}$ and $\ket{1}$, respectively. Although the real quantum computations are prone to small margins of errors, our experiments (see Tables~\ref{tab:simulationpercentage} and \ref{tab:realpercentage})
support the medium-free secure quantum communication approach permissible under QRE interpretation.

\noindent{\bf Contribution.} 
\begin{enumerate}
\setlength\itemsep{1.1em}

\item 
A departing view on the present limits of knowledge about quantum interactions and entanglement is proposed. Evidence and theory supporting our \textit{Quantum Remote Entanglement} (QRE) interpretation suggest that every qubit represented by a quantum particle has distinct probability amplitude values $(\alpha, \beta)$ up to a precision, serving as their identity $QuID$, which could be a foundation for remote entanglement.

\item 
We introduce a medium-free secure quantum communication approach based on the QRE view; this proposed approach potentially sidesteps the challenges of a physical quantum information relay network. 

\item We run actual quantum operations and simulations
on a quantum circuit to support our proposed approach. Quantum information prepared in various states was communicated from a sender to a receiver, indicating a medium-free secure communication possibility. 

\item 
Our work serves as the foundation for future medium-free quantum communication when real-world experiments subsequently verify the herein proposed QRE interpretation.

\end{enumerate}


\vspace{3mm} 
This work does not raise any ethical issues.
    \begin{figure*}[htbp]
    \centering
    \resizebox{.95\linewidth}{!}{%
        \begin{minipage}[b]{.5\textwidth}
        \centering
        \subfloat[][Physical layer fiber optical network with multiple relay nodes in the middle]{
            \includegraphics[width=\linewidth]{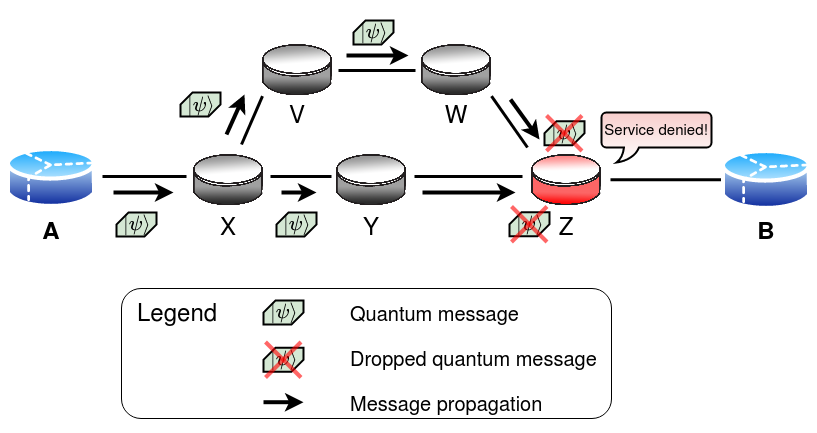}
            \label{fig:sub:teleport_satellite_a}}
        \end{minipage} \hspace{5em} \qquad
        \begin{minipage}[b]{.5\textwidth}
        \centering
            \subfloat[][Space-to-ground network for sending and receiving quantum information via a satellite relay]{
            \includegraphics[width=\linewidth]{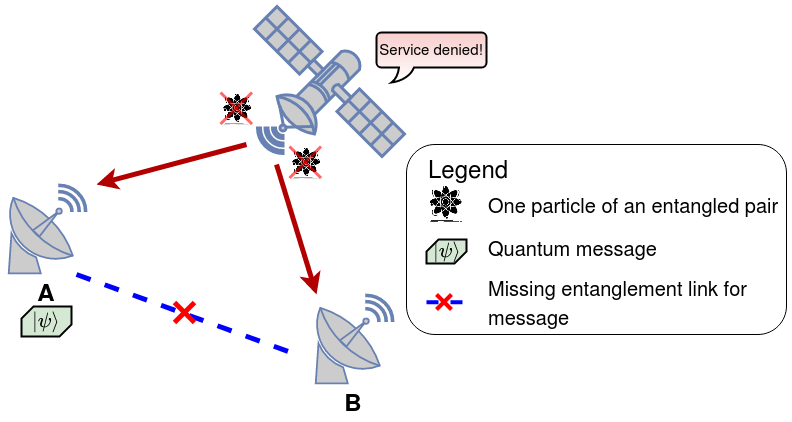}
            \label{fig:sub:teleport_satellite_b}}
        \end{minipage}
    }
    \caption{The two main types of existing quantum communication networks, which transmit information through physical mediums, relying on trusted relays such as quantum repeaters and processors for secure transmission over long distances. 
    }
    \label{fig:teleport_satellite}
    \end{figure*}
    
\section{Fundamentals and Security Model} 
\label{sec:background}
This section gives an overview of secure communication in computer and quantum networks. Next, we discuss the challenges of physical layer security in such systems.
Finally, we consider the security model and potential threats.

\subsection{Secure Communication}
Secure communication between two parties occurs when they are interacting, and any third party is unable to listen in or affect the communication~\cite{kurosecomputer}. It generally prevents malicious actors from eavesdropping or intercepting private communication. 
%
Most popular and widely adopted modern methods to secure data focus on encryption, making it sufficiently robust against unauthorized parties from accessing the data at the higher layers. 
Encryption~\cite{kessler2003overview} is the process of encoding information secured by cryptography. It converts original messages, also known as plaintext $m$, into an encrypted form, known as ciphertext $c$. By communicating through ciphertext $c$, the goal is that only authorized entities are able to decipher a ciphertext to derive intelligible content~\cite{HASSAN2017133}, thereby denying a would-be attacker access to the original information. 
These standard encryption protocols are usually implemented at higher layers of the network stack. However, building security on top of an insecure foundation exposes the communication to threats at lower layers, such as the physical layer (see Section~\ref{sec:related}).

\subsection{Secure Quantum Communication}
Quantum communication networks depend on the quantum mechanical properties of photons instead of classical public or private key cryptography that can be computationally cracked, resulting in more secure networks. However, a significant limitation is that quantum networks do not scale well.

\vspace{3mm} 
\noindent{\bf Quantum Key Distribution (QKD).} 
Quantum cryptography elegantly circumvents this limitation by sending encrypted data as classical bits over networks while securing the messages using keys encoded in quantum states using photons. The photons are used to represent qubits and transmitted as cryptographic keys to decrypt the sent information; hence often referenced as Quantum Key Distribution~\cite{RevModPhys.74.145,RevModPhys.81.1301,Lo_2014,Diamanti_2016,Du_ek_2006,lo2008quantum}. 
The security of encrypted data is ensured by the no-cloning theorem~\cite{Wootters1982Single,DIEKS1982271}, which forbids the perfect reproduction, or cloning, of a quantum state without disturbing it, therefore enabling honest parties to detect the presence of a potential attacker. However, QKD can only produce and distribute a key and not transmit any other messages. It relies on an authenticated classical channel of communication, which is a major drawback, as classical cryptography can achieve secure communications at a fraction of the cost with such a classical channel~\cite{quantum2010007}.

\subsection{Physical Layer Security}
In modern-day communication networks, fiber optics are the standard for telecommunication as they have high bandwidth, reaching data rates in excess of 160 Gb/s~\cite{rashed2017suitable}. Moreover, entangled photons can be sent over standard fiber optics~\cite{Inagaki_2013} for High Photon Efficiency (HPE) optical communications. It presents extraordinary security challenges and vulnerabilities. 
Unfortunately, many attacks target the lowest layer of optical networks, making the physical layer of an optical network vulnerable. Some identified threats are classified into these few categories~\cite{5749282}: 
(Confidentiality) where an adversary tries to listen in on private communications. (Integrity) where an entity alters or manipulates communication messages. (Availability) where an active attacker tries to subvert the successful delivery of messages. (Authentication) where an unauthorized actor tries to communicate as an authorized entity. (Privacy) where an adversary is observing the existence of communications, exposing communication to privacy risks.

Fiber optical networks (see Figure~\ref{fig:sub:teleport_satellite_a}) channeling communication through repeaters rely on trusted links. Any secure communication along the chain at the optical layer must be able to safeguard against exploitation that takes advantage of both direct and relay links for physical layer security, making it an extremely challenging problem for achieving security~\cite{7407417}.
On the ground, fiber optic cables can transport short-lived photon entanglement over short distances, approximately a few hundred kilometers~\cite{10.1145/3341302.3342070}. 
Recent developments in satellite-based networks (see Figure~\ref{fig:sub:teleport_satellite_b}) have achieved the transmission of photons from the satellite to the ground over a great distance~\cite{Liao_2017}.
While space-to-ground links for quantum-state transmission do not rely on multiple trusted relays, it depends on the satellite's physical capabilities, which is a costly single point of weakness.

\subsection{Security Aspect}
The objective of QRE is to (1) submit secure communication through the non-locality nature of quantum theory governing quantum entanglement and (2) overcome the challenges of existing communication networks that rely on physical trusted relays. We introduce medium-free remote entanglement for settings where two parties communicating highly sensitive information cannot trust the intermediate relays in the communication network. In particular, our proposal is based on the view that the inherently non-local nature of quantum theory allows any two specific quantum particles to establish remote entanglement, transmit information, and bypass the need for a physical transmission medium.

\vspace{3mm} 
\noindent{\bf Threat Model.} We assume that an attacker has compromised at least one of the relay nodes along the communication line (e.g., quantum repeaters in a ground optical network, satellite relay in a space-to-ground network), and they control the nodes.
In these scenarios (see Figure~\ref{fig:teleport_satellite}), the attacker could potentially perform denial-of-service attacks, compromise privacy based on metadata leakage, and interfere or disturb transmission signals. Additionally, our threat model centers on attackers who undermine the physical infrastructure of quantum communication networks, reflecting a host of known attacks such as photon-number-splitting~\cite{brassard2000limitations}, time-shift~\cite{qi2006time}, and various other~\cite{xu2020secure,PhysRevA.84.062308,PhysRevA.73.022320,lydersen2010hacking,Lo_2014} attacks.

\section{Mediated Remote Entanglement} %
\label{sec:mediated} 
The entanglement of quantum particles can be created in numerous ways. So far, there are at least two known methods of creating such entangled quantum states. 
In the first method, entanglement is created by direct interactions between component subatomic particles. These components are united initially, such as an Einstein-Podolsky-Rosen (EPR) pair, where the components are parts of a pair of qubits in a Bell state before they are then separated (see Section~\ref{q-entangle}). 
Although there are many possible ways to create entangled Bell states, one of the simplest methods is to perform computations on a quantum circuit. The resulting Bell states are four specific maximally entangled quantum states of two qubits. After which, each part of the pair is then separated for computation or communication purposes. 

The second method, the mediated entanglement (see Section~\ref{mediated}), makes simultaneous quantum measurements on spatially separated non-entangled qubits. 
By coherently performing the measurements (e.g., via laser or micro-wave pulsations), it unifies the separate qubits into a single system, even though they are initially spatially separated. As a result, the components are condensed (Bose-Einstein condensation), where every Bose-Einstein condensate is in a highly entangled state because the particles in a condensate are coherently distributed over space~\cite{PhysRevA.66.052323}. 
This technique has been used in superconducting circuits~\cite{narla2016robust,campagne2018deterministic}, quantum communication networks~\cite{humphreys2018deterministic,leung2019deterministic}, and other quantum systems~\cite{bienfait2019phonon}, showing much potential for quantum information technology.

Effective quantum communication between remote quantum nodes, requiring high fidelity quantum state transfer and remote entanglement generation, has been proposed using the second method~\cite{leung2019deterministic,chang2020remote}. 
Such remote entanglement has been previously realized using various probabilistic schemes. However, recent deterministic remote entanglement schemes using a variety of superconducting circuit approaches have been successfully demonstrated. This deterministic entanglement can significantly minimize transmission channel loss, achieve high entanglement fidelities, and build large-scale quantum communication systems.

\subsection{Quantum Entanglement}
\label{q-entangle}

One fundamental property of quantum mechanics and an essential concept in quantum information science is the entanglement, in which separate qubits unify to become a combined quantum system. 
This unified system is governed by one quantum wave function. The quantum states of each qubit in this entangled system can only be described with reference to other qubits. Even though the qubits might be spatially apart, the joint quantum-mechanical measurement of entangled qubits results in correlations between physical properties of the system that are observable. These quantum correlations are stronger than classical correlations. In such a unified system, the measurement of one qubit may inadvertently ``influence'' the other qubits entangled with it. 

In 1935, the Einstein-Podolsky-Rosen paradox~\cite{einstein1935can} (EPR paradox) was a thought experiment proposed by physicists Albert Einstein, Boris Podolsky, and Nathan Rosen. 
They argued that quantum mechanics was an incomplete physical theory and designed a thought experiment to disprove entanglement.
%
However, in 1983, one such experiment was performed using photon pairs. Aspect et al.~\cite{aspect1982experimental} experimentally realized the EPR thought experiment, demonstrating that correlated measurements of the photon spin components along arbitrary directions resulted in a complementary instantaneous reduction. They used two separated detectors (two-channel polarizers such as optical analogs of Stern-Gerlach filters). Similar experiments have been repeated over larger distances by sending polarized photons through fiber optic cables, yielding the same conclusive results~\cite{tittel1998experimental}. Thus, even though we have yet to figure out the faster-than-light transmission of information via entanglement, it is widely implemented in quantum key distribution that performs cryptographic protocols~\cite{RevModPhys.74.145,Du_ek_2006} to secure communication.

\subsection{Mediated Entanglement}
\label{mediated}
The mediated entanglement, another form of entanglement which realizes remote entanglement, occurs in quantum-coherent systems. One such system was proposed in 1924 by Bose and Einstein~\cite{bose1924plancks}. The Bose-Einstein condensate is a state of matter in which separate subatomic particles are cooled to near absolute zero (0 Kelvin), where zero reflects the complete absence of thermal energy. At this point, the separate particles fuse into a single quantum mechanical system that a single wave function can describe. If any particle in the system is perturbed, the rest of the system, following quantum-mechanical laws, is affected and reacts accordingly. Only in 1995, the first Bose-Einstein condensates were produced in a vapor of rubidium-87 atoms to form gaseous condensates~\cite{Anderson198}. Subsequently, cesium atoms forming Bose-Einstein condensates have exhibited entanglement among trillions of component atoms~\cite{julsgaard2001experimental}.

Existing mediated entanglement methods can be broadly categorized into two stages of development. In the first stage, the remote entanglement generation is based on probabilistic schemes~\cite{Ying_Qiao_2005,narla2016robust}. These schemes tend to apply entangling measurements in the microwave domain that probabilistically project unentangled superposition states onto entangled states. The second stage takes a deterministic approach~\cite{humphreys2018deterministic,leung2019deterministic} for the generation of entangled Bell states with high fidelity rates. This approach requires the rate of entanglement generation to exceed losses in the transmission line and decoherence of each qubit.
Both groups aim to generate these remote entangled pairs, entangling two remote quantum systems that never interact directly.
Several recent experiments have demonstrated both probabilistic and deterministic remote entanglement generation between superconducting qubits, with 60--95\% entanglement fidelities~\cite{zhong2019violating,kurpiers2018deterministic,dickel2018chip}.

\vspace{3mm} 
\noindent{\bf Probabilistic.} 
In certain cases, remote entanglement generation follows a statistical model. Some cases~\cite{PhysRevLett.112.170501,dickel2018chip} design measurements on a quantum system that purifies quantum correlations of an entangled state to induce entanglement. 
Two superconducting qubits, in a particular case, coupled to the same microwave resonator have been entangled using such a measurement~\cite{PhysRevLett.112.170501}. The qubits are separated by 1.3 meters. By engineering a continuous measurement where one of the three outcomes is a Bell state, they achieved probabilistic measurement-induced entanglement generation.
In another case, remote entanglement was generated using flying, single photons~\cite{narla2016robust}. The remote entanglement experiment was performed with a single-photon detector based on a superconducting qubit, where the flying photon is robust to transmission losses. It offers the advantage that the production of pure entangled states depends on the probability of the successful detection of photons.
While these probabilistic strategies have been widely used to realize high-fidelity remote entanglement, their probabilistic nature, sometimes albeit with low success probability, limits the rate of communication.

\vspace{3mm} 
\noindent{\bf Deterministic.} 
As quantum information communication systems grow from two qubits to large-scale networks, the rate of communication undoubtedly increases. It requires the scaling of the entanglement of distant systems that do not interact directly. Naturally, deterministic entanglement generation between distant qubits is required to support such rates of communication. 
One favored approach to achieve the deterministic entanglement generation~\cite{axline2018demand,kurpiers2018deterministic,leung2019deterministic,zhong2019violating,zhong2019violating,chang2020remote} is through photonic transfer via transmission lines such as coaxial cables. 
Leung et al.~\cite{leung2019deterministic} established a bidirectional photonic communication channel between two qubits. The required coaxial cable connection allows the photons to be transferred coherently through the discrete modes of the channel, avoiding any loss caused by external factors that severely limit the communication ﬁdelity.
Axline et al.~\cite{axline2018demand} implemented a deterministic state transfer protocol by employing superconducting microwave cavities, serving as remote quantum memory endpoints in a simple network. Using this memory in the communication modes, they are strongly coupled to a transmission line to realize deterministic entanglement for communication.

Another approach uses microwave components for the deterministic generation of entanglement. It requires an efficient absorption by one photon of the electromagnetic field emitted by the other photon, resulting in a desired propagation of information through a network. For instance, a scheme reporting the deterministic generation of two distant superconducting qubits employs microwave pumps to concurrently and coherently excite both qubits in a buffer resonator~\cite{campagne2018deterministic}. Due to the stimulation, one of the qubits then leaks out and travels through microwave components, and it is captured by a third qubit with a similar scheme. In another protocol~\cite{humphreys2018deterministic}, microwave pulses are used to create entangled states. The deterministic entanglement is achieved using fully heralded single-photon entanglement. 
When the rate of entanglement generation between nodes exceeds the decoherence (loss) rate of entanglement, intrinsically probabilistic entangling protocols can provide deterministic remote entanglement at pre-specified times. Entanglement generation is attempted until success; then, microwave pulses are applied to rotate and create the desired state coherently.

    \begin{figure*}[htbp]
        \centering
        \includegraphics[width=.7\linewidth]{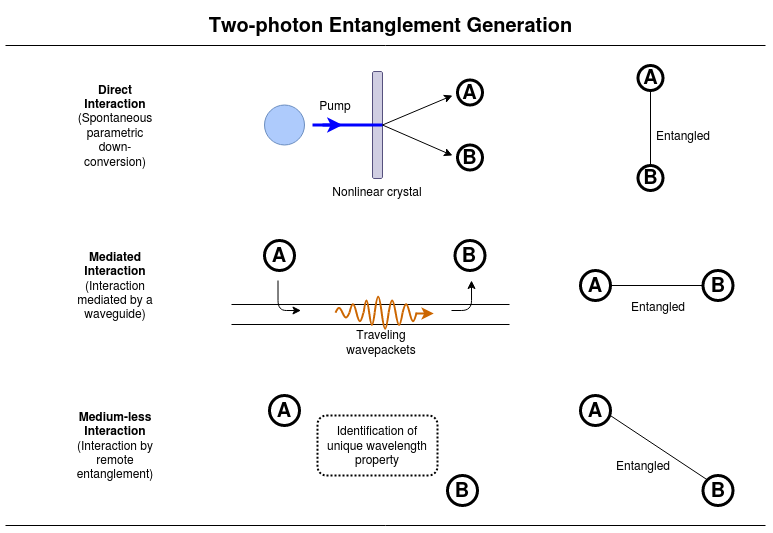}
        \caption{Some of the approaches to generate the quantum entanglement of two particles, $A$ and $B$. 
        The first approach creates entanglement through direct interactions between subatomic particles, converting one photon of higher energy to a pair of photons with lower energy (top). 
        The second approach performs mediated entanglement by using wavepackets to remotely entangle separate qubits (middle). 
        Our approach suggests medium-less entanglement of remote particles by using the distinct identifiers of each qubit (bottom).
        }
        \label{fig:supporting}
    \end{figure*}        

Last but not least, phonons, the particles of sound, have been proposed as a universal medium for coupling remote quantum systems~\cite{PhysRevB.62.8410,schutz2017universal}. The remote entanglement of superconducting qubits has also been demonstrated using phonon-mediated communication. In particular, surface acoustic wave (SAW) phonons are proposed to realize the coherent transfer of quantum states between two superconducting qubits~\cite{bienfait2019phonon}. A single superconducting qubit, launching a roaming phonon into a SAW resonator, allows the phonon to be completely injected into the acoustic channel before re-exciting the emitting qubit. Next, this emitting qubit can recapture the phonon later and perform remote qubit entanglement with high fidelity.

\section{Quantum Remote Entanglement} 
\label{sec:formulate}   
In this section, we present our view on remote entanglement. Imagine the methods to create entanglement of two qubits (direct, mediated, and medium-less interactions alike; see Figure~\ref{fig:supporting}). The main idea of our interpretation lies in the distinguishability of each quantum particle. In this interpretation, every quantum particle is defined by its quantum state with an intrinsic particle wavelength property, making it identifiable. We define this inherent property as a $QuID$ and introduce our approach to a medium-free and secure quantum communication. We formally state the proposed idea in detail below.


\subsection{Proposal Formulation} 
\label{formulate}

We focus on the novel proposal of achieving medium-free quantum communication. 
Let $\mathcal{G}_{\Phi} = (\Phi, \mathcal{E})$ denote the universal quantum body with particles $\phi_i \in \Phi$ and entanglements $(\phi_i, \phi_j) \in \mathcal{E}$. Each particle $\phi_i$ has a property, an intrinsic wavelength, and it is distinct and identifiable, associated with its identity information. 
Using a fundamental quantum mechanical phenomenon, the quantum entanglement, any two particles in the universal quantum body $\mathcal{G}_{\Phi}$ can entangle using their identities. 
Consequently, the two particles freely transmit quantum information using the entanglement as a resource and following a teleportation protocol, thereby achieving medium-free secure communication.

We propose a progressive view of quantum interactions. 
In this view, we suggest that the fundamental mechanics of quantum entanglement, consistent with the non-locality nature of quantum theory, is not conditional on the physical proximity of particles. Instead, we theorize that every quantum particle is distinct in the universal quantum body $\mathcal{G}_{\Phi}$. Each particle has a characteristic wavelength with infinite precision, defined as a quantum identity $QuID$. They can remotely interact with each other from any distance using their identities and transmit quantum information. 
We formally state our view below.

\begin{interpret}[Quantum Remote Entanglement (QRE)]
\label{interpre:QRE}
We postulate that every particle $\phi_i$ in the universal quantum body $\mathcal{G}_{\Phi}$ has a distinct particle wavelength with infinite precision. 
An elementary particle, the photon, quantum of the electromagnetic field, represents a quantum bit of information, qubit $\ket{\psi} = \alpha \ket{0} + \beta \ket{1}$, where $\alpha, \beta \in \mathbb{C}$ are complex numbers with real continuous values of infinite precision.
Qubits are distinctly identifiable by their identity. We define this identity property as $QuID (i) = (\alpha_i, \beta_i) \in \mathcal{C}$.
Quantum particles freely entangle and interact from any physical distance using their $QuID$s. 
\end{interpret}

Moreover, the no-cloning theorem~\cite{DIEKS1982271,Wootters1982Single}, a central principle of quantum information theory that forbids the creation of identical copies of quantum states, states that it is impossible to create independent and duplicate copies of an arbitrary unknown quantum state. 
While it does not prevent one from having several qubits in the same state, this can only happen when the coefficients $(\alpha, \beta)$ are provided. Any given values are an approximation. 
This theorem has profound implications for quantum communication; it is impossible to make a perfect copy of some quantum information. 
Perhaps it could be that the no-cloning theorem also implies QRE.

\vspace{3mm} 
\noindent{\bf Thought experiment.} 
To see this, consider the following thought experiment, and let us suppose that particles and quantum states could be cloned to an exact replication. In addition, we assume Alice and Bob each hold one part of a maximally entangled Bell state. Using this entangled pair as a resource, Alice could now send a particle across vast space to Bob while holding on to the same particle on her end, which is contradictory in itself.
Such a process would result in one particle physically existing in two places simultaneously (possibly violating the no-cloning theorem~\cite{DIEKS1982271,Wootters1982Single} for pure states and its corollary, the no-broadcasting theorem~\cite{Barnum_1996} for mixed states). 
Therefore, the following consequence provides further evidence that suggests that each quantum particle is distinct.

\subsection{Interpretation} 
Now, given that $\phi_i$ knows the identity $QuID(j)$ of $\phi_j$, which can be reliably determined using quantum tomography~\cite{lvovsky2009continuous}, it wants to entangle with $\phi_j$. By the mechanical property of quantum bodies, $\phi_i$ entangles with $\phi_j$ to form an entangled pair $\phi_{ij}$. When the entanglement is established, the sender $\phi_i$ can then employ a teleportation protocol and transmit quantum information from one location to receiver $\phi_j$ any distance away. This process depends solely on the properties of quantum mechanics; no physical medium is involved, thereby achieving a medium-free channel for secure communication. 

In this work, we propose a medium-free quantum communication approach. 
By suggesting that each particle $\phi_i$ is distinctly identifiable by its $QuID(i)$, it enables us to entangle any particles using their $QuID$s and establish communication channels.
Introducing the \textit{Medium-Free Quantum Communication} proposal here, defined as: 
\begin{center}
\noindent\parbox{0.9\linewidth}{%
\begin{define}
    Suppose two quantum particles, $A$ and $B$, that do not currently have a communication channel, want to establish a connection for secure communication, where $A$ conveys some quantum information $\ket{\psi}$ to $B$.
    Given that $A$ has the $QuID$ of $B$, $A$ can directly entangle with $B$ using $QuID(B)$. Now both of them are connected. This newly formed entanglement allows $A$ to transmit $\ket{\psi}$ to $B$, secured by the properties of quantum mechanics, through any communication means such as the quantum teleportation protocol, thereby achieving a medium-free secure communication channel.
\end{define}
}%
\end{center} 
Following this proposal, it seems that we could leverage the distinct quantum identities, valuable in establishing quantum entanglements across vast distances, to establish a secure communication channel between particles without the need for a physical medium or intermediary to transport them.


\section{Can Quantum Communication be Medium-Free?}
\label{sec:approach}
A different approach is proposed here that supports medium-free secure quantum communication.
First, it requires the distinct $QuID$ of any qubit to be entangled. Second, the quantum effort appears to grow linearly with the number of qubits. Third, it can be used to perform remote entanglement from any distance.
By supposing a universal quantum body, 
each qubit represented by a particle can be written as: 

\begin{equation}
    \ket{\psi}_i = \alpha_i \ket{0} + \beta_i \ket{1}
\end{equation}
with $\lvert \alpha_i \rvert + \lvert \beta_i \rvert = 1$.
Our view is that both values, $\alpha_i$ and $\beta_i$, defining the qubit state, have continuous values of infinite precision, giving it an identity. We define this identity as a $QuID(i) = (\alpha_i, \beta_i)$, and any qubit can entangle with another using this $QuID$ to perform the formulated secure communication. Hence, quantum entanglement is not contingent on physical proximity. Any qubit can entangle with another qubit regardless of the distance between them.

For now, assuming that the \textit{Quantum Remote Entanglement} interpretation holds, it allows two communicating parties to transmit information in quantum states. The fundamental mechanics of quantum nature is that the act of measuring a quantum system disturbs the system. This results in an essential property of quantum communication, where two parties' communication has the ability to detect the presence of an eavesdropper trying to gain information of their communication, thereby securing the communication channel.  

Following the canonical quantum teleportation protocol~\cite{PhysRevLett.70.1895}, we can obtain secure quantum communication between two parties. For example, suppose Alice has a qubit $\ket{\psi} = \alpha \ket{0} + \beta \ket{1}$ in a prepared state that she wants to send to Bob. The resources required for quantum teleportation are (1) a communication channel capable of transmitting two classical bits, (2) a Bell measurement on one of the EPR pair qubits, and (3) the quantum state manipulation of the other qubit in the pair. 

The protocol is as follows. 
Given that Alice has a qubit $A$, Bob has a qubit $B$, and Alice has the knowledge of Bob's qubit identity $QuID(B)$, she entangles her qubit $A$ with Bob's qubit $B$ through its $QuID$, using QRE, or possibly any other implementation of remote quantum operations~\cite{huelga2001quantum,vishnu2018experimental,lv2018joint} upon a distant quantum system (e.g., by local operations, classical communication); thereby generating an entangled pair, defined by:

\begin{equation} 
    \ket{\phi^+} = \frac{1}{\sqrt{2}}(\ket{0}_A \otimes \ket{0}_B + \ket{1}_A \otimes \ket{1}_B)  
\end{equation}
where Alice ($A$) and Bob ($B$) each possess one qubit of the entangled pair, respectively.

    \begin{figure*}[!htbp] 
        \centering
        \includegraphics[width=.8\linewidth]{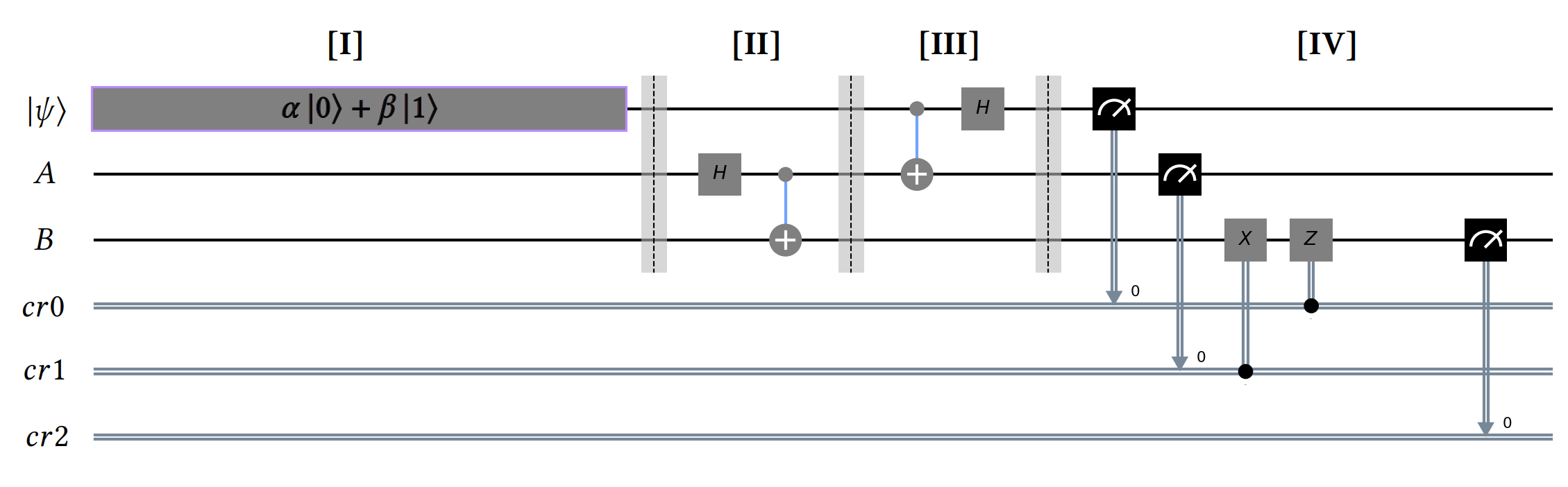}
        \caption{Quantum circuit to demonstrate our approach to medium-free secure communication. The first three solid single lines are quantum registers in the circuit representing (1) information $\ket{\psi}$, (2) Alice's qubit $(A)$ and Bob's qubit $(B)$. The next three double lines are classical registers to store measurements from the quantum registers.
        }
        \label{fig:fullcircuit}
    \end{figure*}        

When Alice prepares her qubit $\ket{\psi}$ that she needs to transmit in a prepared state and entangles $\ket{\psi}$ with her qubit $A$ in the EPR pair $\ket{\phi^+}$, it creates a three-qubit quantum system, resulting in the starting state:

\begin{equation} \label{eq:threequbit}
\begin{split}
    \ket{\psi} \otimes & \ket{\phi^+} = \\
    &= \frac{1}{\sqrt{2}}(\alpha \ket{0} \otimes (\ket{00} + \ket{11}) + \beta \ket{1} \otimes (\ket{00} + \ket{11}))  \\
    &= \frac{1}{\sqrt{2}}(\alpha \ket{000} + \alpha \ket{011} + \beta \ket{100} + \beta \ket{111})  \\
\end{split}
\end{equation}
Next, Alice applies a CNOT gate on her qubits $\ket{\psi}$ and $A$, which transforms the state to: 

\begin{equation} \label{eq:steptwoCNOT}
\begin{split}
    (&CNOT \otimes I) (\ket{\psi} \otimes \ket{\phi^+}) = \\
    &= (CNOT \otimes I) \frac{1}{\sqrt{2}}(\alpha \ket{000} + \alpha \ket{011} + \beta \ket{100} + \beta \ket{111}) \\
    &= \frac{1}{\sqrt{2}}(\alpha \ket{000} + \alpha \ket{011} + \beta \ket{110} + \beta \ket{101})
\end{split}
\end{equation}
followed by a Hadamard transform to the first qubit state $\ket{\psi}$, resulting in:

\begin{equation} \label{eq:steptwoH}
\begin{split}
    (H \otimes I \otimes I) & \frac{1}{\sqrt{2}} (\alpha \ket{000} + \alpha \ket{011} + \beta \ket{110} + \beta \ket{101}) = \\
    &= \frac{1}{\sqrt{2}}(\alpha(\ket{000} + \ket{011} + \ket{100} + \ket{111}) 
        + \\
        & \hspace{5mm} + \beta(\ket{010} + \ket{001} - \ket{110} - \ket{101}))
\end{split}
\end{equation}
Then, Alice performs a Bell measurement of her EPR pair, qubits $A$ and $\ket{\psi}$. This collapses Alice's qubits into binary values, yielding one of four measurement outcomes with an equal probability of $\frac{1}{4}$. 
The outcome results are encoded in two classical bits of information and sent to Bob, and $B$'s state will be projected to the following states:
\begin{enumerate}
\setlength\itemsep{0.5em}
    \item \textbf{Alice measures 00}, then $\ket{00} \rightarrow \bigg( \alpha \ket{0} + \beta \ket{1} \bigg)$
    \item \textbf{Alice measures 01}, then $\ket{01} \rightarrow \bigg( \alpha \ket{1} + \beta \ket{0} \bigg)$ 
    \item \textbf{Alice measures 10}, then $\ket{10} \rightarrow \bigg( \alpha \ket{0} - \beta \ket{1} \bigg)$ 
    \item \textbf{Alice measures 11}, then $\ket{11} \rightarrow \bigg( \alpha \ket{1} - \beta \ket{0} \bigg)$ 
\end{enumerate}
Once Bob receives the two classical bits from Alice, he modifies his EPR pair qubit accordingly, depending on the results. The appropriate unitary operation(s) that he applies are based on the following: $00$ (Identity gate), $01$ (X-gate), $10$ (Z-gate), $11$ (Z-gate followed by X-gate). 
The resulting qubit is identical to Alice's $\ket{\psi} = \alpha \ket{0} + \beta \ket{1}$. 

Hence, assuming the Quantum Remote Entanglement interpretation, where particles with distinct identities entangle with other qubits using their QuIDs, quantum entanglement is now freed of dependence on physical proximity. Next, following a teleportation protocol, our proposed approach completes the transmission of quantum information for medium-free secure quantum communication. 



\section{Quantum Circuit Experiments}
\label{sec:experiments}
In this section, we present experimental results, where Alice securely sends quantum information by transferring her state $\ket{\psi}$ to Bob. It provides a proof of concept to demonstrate the feasibility of our proposal, validating the approach to achieve medium-free secure communication. The purpose is to establish that our proposal has practical considerations for future quantum communication. 

Following the Qiskit quantum circuit~\cite{Qiskit-Textbook} for teleportation, we demonstrate the potential of our approach 
by testing the circuit with its in-built simulators and running the circuit on a real noisy intermediate-scale quantum computer.
We run all the experiments using Qiskit 0.12.0, an open-source quantum computing framework that supports Python 3.6.9. While the simulations are on a Linux server with 128GB of RAM and a 32-core processor, the actual quantum computations are performed on cloud-based quantum computing services by the IBM Quantum provider (\texttt{hub=`ibm-q'}).

\subsection{Quantum Circuit Setup}
\label{subsec:circuit}
We build a quantum circuit that is a widely used model for quantum computation, consisting of circuit wires to represent qubits and classical bits, and boxes to represent quantum operations. The quantum gates are reversible transformations. It mainly provides us with an architecture for formulating the physical construction of quantum computers.
While this quantum circuit setup requires the $A$ and $B$ pair to be prepared in Bell states and not any arbitrary states, it allows us to demonstrate the secure transmission of information between two qubits (see Figure~\ref{fig:fullcircuit} for the detailed circuit stages I--IV and gate operations).

Suppose Alice has a qubit $A$, and she wants to send some prepared quantum information $\ket{\psi} = \alpha \ket{0} + \beta \ket{1}$ to Bob, who has another qubit $B$. Now, assume that QRE (Interpretation~\ref{interpre:QRE}) holds, and particles represented by qubits can entangle from any physical distance using their distinct identity $QuID$.
Alice, who knows $QuID$ of $B$, first entangles her qubit $A$ with $B$ only using the $QuID(B)$. She can then initiate the teleportation protocol, using the newly entangled pair as a resource to securely send $\ket{\psi}$ to Bob, thereby completing the medium-free remote transmission of quantum information. 
This is achieved in the four stages of the circuit by:

\begin{enumerate}[label = {[Stage \Roman*]},align=left]
\setlength\itemsep{0.5em}
    \item Alice first prepares the information to be communicated in the quantum state $\ket{\psi}$.

    \item Next, a Hadamard (\textit{H}) gate is applied to $A$, and a Controlled NOT (\textit{CNOT}) gate is applied onto $B$ controlled by $A$.

    \item At this point, Alice has two qubits, $A$ and $\ket{\psi}$, while Bob has one qubit, $B$. The three-qubit system state is when Alice applies a \textit{CNOT} gate to $A$, controlled by $\ket{\psi}$, and another \textit{H} gate to $\ket{\psi}$. 
    
    \item Alice measures both of her qubits, $A$ and $\ket{\psi}$, storing the results in two classical bits and sending them to Bob. Upon receiving the two classical bits, Bob applies the \textit{X} or \textit{Z} gates accordingly.
\end{enumerate}
At the end of this quantum circuit setup, Alice's qubit is teleported to Bob. It results in Bob with $\ket{\psi}$, as he successfully reconstructs Alice's qubit state and obtains the communication in the exact quantum state.

\subsection{State Vector Simulation}

The circuit specified in Section~\ref{subsec:circuit} allows us to simulate the communication of a qubit in quantum state $\ket{\psi}$. However, we are currently unable to specify a two-level quantum mechanical system exactly as $\alpha$ and $\beta$ as complex numbers. By confining $\alpha$ and $\beta$ to real numbers and adding a relative phase term, we can describe the restricted qubit state. Using such restriction, we plot visual representations of a state with a Bloch sphere~\cite{PhysRev.70.460}, geometrically representing the pure state space of a qubit.

\begin{figure}[htbp]
        \centering
        \includegraphics[width=.35\linewidth]{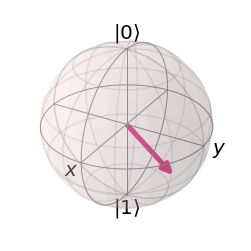}
        \caption{Bloch sphere representation of the quantum state $\ket{\psi}$ (Equation~\ref{eq:statevec_0}) to be communicated from Alice to Bob.}
        \label{fig:statevec_0}
    \end{figure}

In this state vector simulation, which allows a perfect single-shot execution of quantum circuits and returns the final state vector of the simulation, Alice first prepares the restricted state $\ket{\psi}$ as:
\begin{equation} \label{eq:statevec_0}
\ket{\psi} =  
\begin{pmatrix}
\alpha    \\
\beta   \\
\end{pmatrix} = 
\begin{pmatrix}
-0.57659 + 0.24170i    \\
-0.59478 - 0.50532i   \\
\end{pmatrix}
\end{equation}
which we visually present the state $\ket{\psi}$ by plotting it in a Bloch sphere representation below (see Figure~\ref{fig:statevec_0}).

\begin{figure}[htbp]
        \centering
        \includegraphics[width=1.0\linewidth]{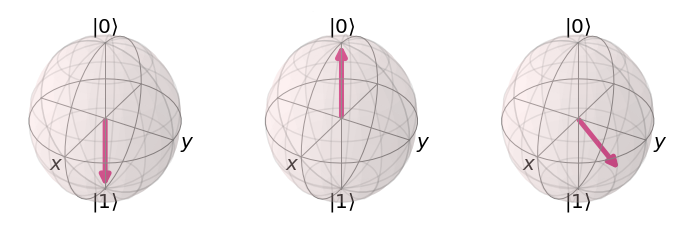}
        \caption{At the end of the circuit, both of Alice's qubits, $\ket{\psi}$ (left) and $A$ (middle), collapse to either $\ket{0}$ or $\ket{1}$, and Bob's qubit $B$ (right) is the same as the initially prepared communication state $\ket{\psi}$.}
        \label{fig:statevec_2}
    \end{figure}

\noindent Next, we run the simulator on the circuit, where various gates and measurements are applied to the quantum and classical registers. It simulates the teleportation protocol and transmits the state $\ket{\psi}$ from the first quantum register to the third quantum register. 
In the end, the quantum circuit outputs a three-qubit state vector as below:

        \begin{equation*}
        \substack{\ket{\psi} \\ \text{output} \\ \text{state vector}} = 
        \begin{bmatrix}
        0 \\
        -0.57659+0.24171i \\
        0 \\
        0 \\
        0 \\
        - 0.59478 - 0.50532i \\
        0 \\
        0 \\
        \end{bmatrix}
        \end{equation*}
where both of Alice's qubits, $\ket{\psi}$ (left) and $A$ (middle), collapse to either $\ket{0}$ or $\ket{1}$, while Bob's qubit $B$ (right) becomes the $\ket{\psi}$ as prepared by Alice prior to the transmission, where $(\alpha_B,\beta_B) = (\alpha_{\ket{\psi}},\beta_{\ket{\psi}}) = (-0.57659 + 0.24170i, -0.59478 - 0.50532i)$; $\ket{\psi}$ has been successfully teleported from Alice to Bob, completing the quantum communication (see Figure~\ref{fig:statevec_2}). 

In addition, we run the circuit multiple times with the prepared communication qubit $\ket{\psi}$ in various different states. We notice that in the results, the first two qubits owned by Alice either end up in $\ket{0}$ or $\ket{1}$ every time, but the third qubit held by Bob is always the same as $\ket{\psi}$, in the initially prepared state (see Appendix for the additional simulations).

\subsection{Noisy Quantum Circuit Computation} 
The current state-of-the-art quantum circuit computers do not provide complete fault-tolerant implementations. As a result, the circuit computations are noisy without a full error correction mechanism, and all existing quantum computers fall under this category.
While the existing hardware is not able to sample state vectors, we can demonstrate on a single quantum chip that the operations in the circuit have correctly performed the teleportation of qubit information.

\vspace{3mm} 
\noindent{\bf Simulation.} 
To demonstrate the transmission of a qubit $\ket{\psi}$ from Alice to Bob, we first run a simulation with no computation errors.
Alice first prepares and sets her qubit $\ket{\psi}$ to the state $\ket{0}$. Then applying an $H$-gate to initialize the qubit into a superposition state $\ket{\psi} = \frac{1}{\sqrt{2}} \ket{0} + \frac{1}{\sqrt{2}} \ket{1}$, she has the quantum state that is to be teleported to Bob. 

    \begin{figure}[htbp]
        \centering
        \includegraphics[width=.6\linewidth]{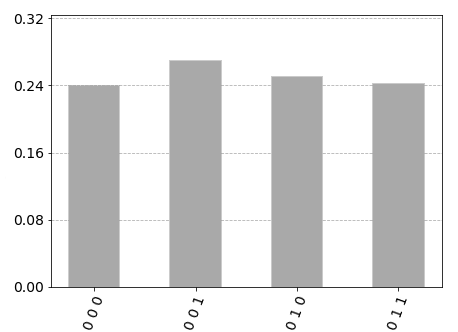}
        \caption{There is a $100\%$ chance of measuring Bob's qubit $B$ in the state $\ket{0}$ (all four of the leftmost bit are zeros), indicating that the circuit successfully transported Alice's $\ket{\psi}$ to Bob.}
        \label{fig:simulateket0}
    \end{figure}        

Since all quantum gates are unitary, making the operations reversible, we can take the inverse of the $H$-gate on $\ket{\psi} = \frac{1}{\sqrt{2}} \ket{0} + \frac{1}{\sqrt{2}} \ket{1}$ to get back to $\ket{0}$. Hence, running and sampling the circuit a number of repetitions, we are able to prove the qubit $\ket{\psi}$ has been successfully communicated from Alice to Bob, as we measure $\ket{0}$ with $100\%$ certainty. 
We sample the circuit $1024$ times to get a frequency distribution of the results. We plot this distribution of measured bits in Figure~\ref{fig:simulateket0}. It can be seen that in the $x$-axis, all four of the leftmost bit are zeros, indicating that the $\ket{\psi}$ prepared by Alice has been successfully transmitted to Bob and the circuit worked properly.

    \begin{figure}[htbp]
        \centering
        \includegraphics[width=.6\linewidth]{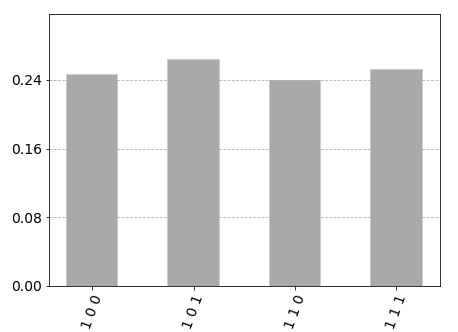}
        \caption{There is a $100\%$ chance of measuring Bob's qubit $B$ in the state $\ket{1}$ all four of the leftmost bit are ones, indicating that when the prepared $\ket{\psi} = \ket{1}$, the circuit successfully transported $\ket{\psi}$ to Bob.}
        \label{fig:simulateket1}
    \end{figure}        

In addition, we repeat the same simulation for $\ket{\psi}$ in state $\ket{1}$. The prepared qubit $\ket{\psi}$ initialized into superposition state $\ket{\psi} = \frac{1}{\sqrt{2}} \ket{0} - \frac{1}{\sqrt{2}} \ket{1}$, and transmitted to Bob. We sample it $1024$ times. Similarly, the measurement results returned $\ket{1}$ with $100\%$ certainty (all four of the leftmost bit in the $x$-axis are ones), indicating that the qubit $\ket{\psi}$ has been successfully communicated from Alice to Bob (see Figure~\ref{fig:simulateket1}).

    \begin{table}[htbp]
    \centering
    \caption{Measurement distribution of all possible results for simulation of $\ket{\psi}$ in both states $\ket{0}$ and $\ket{1}$. The first bit on the left is Bob's measured bit, and the next two bits on the right belong to Alice.}
        \resizebox{.75\linewidth}{!}{%
        \begin{tabular}{|c|c|c|}
        \hline
        \textbf{Qubit} \bm{$\ket{\psi}$} & \textbf{Measurement bits} &\textbf{Results ($\%$)}  \\ \hline
        \multirow{4}{*}{$\ket{0}$}      &000      &26.6    \\ \cline{2-3}
                                        &001      &26.0 \\ \cline{2-3}
                                        &010      &22.9 \\ \cline{2-3}
                                        &011      &24.5 \\ \hline
        \multirow{4}{*}{$\ket{1}$}      &100      &24.2    \\ \cline{2-3}
                                        &101      &26.0 \\ \cline{2-3}
                                        &110      &25.3 \\ \cline{2-3}
                                        &111      &24.5 \\ \hline
        \end{tabular}
        }
    \label{tab:simulationpercentage}
    \end{table}

As shown in Table~\ref{tab:simulationpercentage}, the \textit{Measurement bits} column, while the states of the two rightmost bits either collapse to `0' or `1', the leftmost bit (Bob's measured qubit) always ends up in the same state as the prepared qubit $\ket{\psi}$. If $\ket{\psi} = \ket{0}$, then Bob's measured bit is `0', else if $\ket{\psi} = \ket{1}$, then Bob's measured bit is `1'.

\vspace{3mm} 
\noindent{\bf Real Quantum Computer.} 
Next, we run the quantum circuit, which has been successfully simulated, on a real noisy intermediate-scale quantum computer.
In this real quantum computation, two experiments are performed. First, Alice prepares two qubits to communicate and sets them in the superposition states:
\begin{enumerate}
\setlength\itemsep{0.5em}
    \item \textbf{qubit $\ket{\psi}_a$ in state $\ket{0}$:} 
    \begin{equation*}
        \ket{\psi}_a = \ket{0} \rightarrow \bigg( \frac{1}{\sqrt{2}} \ket{0} + \frac{1}{\sqrt{2}} \ket{1} \bigg)
    \end{equation*}
    \item \textbf{qubit $\ket{\psi}_b$ in state $\ket{1}$:}  
    \begin{equation*}
        \ket{\psi}_b = \ket{1} \rightarrow \bigg( \frac{1}{\sqrt{2}} \ket{0} - \frac{1}{\sqrt{2}} \ket{1} \bigg) 
    \end{equation*}
\end{enumerate}
She now has the quantum states to communicate with Bob. We sample the circuit $1024$ times each for both the communication of $\ket{\psi}_a$ and $\ket{\psi}_b$. The two frequency distributions of the resulting measured bits are plotted below in Figure~\ref{fig:realcompute}. 

    \begin{figure*}[htbp]
    \centering
    \resizebox{.75\linewidth}{!}{%
        \begin{minipage}[b]{.5\textwidth}
        \centering
        \subfloat[][Prepared communication state: $\ket{\psi}_a = \ket{0}$ ]{
            \includegraphics[width=\linewidth]{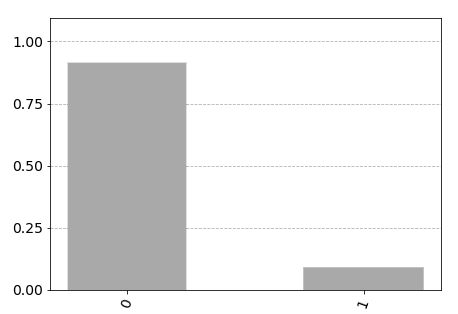}
        \label{fig:realcompute_a}}
        \end{minipage} \hspace{5em} \qquad 
        \begin{minipage}[b]{.5\textwidth}
        \centering
        \subfloat[][Prepared communication state: $\ket{\psi}_b = \ket{1}$ ]{
            \includegraphics[width=\linewidth]{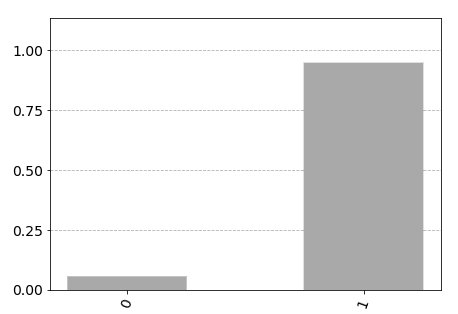}
            \label{fig:realcompute_b}}
        \end{minipage}
    }
    \caption{Theoretically, the measured results should be the initially prepared state $100\%$ of the time, but in a real noisy quantum circuit, a small percentage of the measured bits are errors due to noise in the gates and the qubits. In (a), where $\ket{\psi}_a = \ket{0}$, there is a $0.089$ error rate. In (b), where $\ket{\psi}_b = \ket{1}$, there is a transmission error rate of $0.056$. }
    \label{fig:realcompute}
    \end{figure*}

In Figure~\ref{fig:realcompute_a}, for the transmission of $\ket{\psi}_a$ from Alice to Bob, eventual measurements on Bob's qubit returned the `0' result $933$ times, and the `1' result in the remaining $91$ repetitions, resulting in an experimental error rate of $0.089$. 
As for the transmitted $\ket{\psi}_b$, in Figure~\ref{fig:realcompute_b}, the results on Bob's end returned the `0' measurement $57$ times and the `1' measurement in the remaining $967$ repetitions, achieving an experimental error rate of $0.056$.

    \begin{table}[htbp]
    \centering
    \caption{Real quantum circuit computation results for all possible measurement states of Bob's qubit $(B)$. The circuit returned measurement `0' in 91.1\% of the times when $\ket{\psi}$ is prepared in $\ket{0}$, and measurement `1' in 94.4\% of samples when $\ket{\psi}$ is prepared in $\ket{1}$.}
        \resizebox{.75\linewidth}{!}{%
        \begin{tabular}{|c|c|c|}
        \hline
        \textbf{Qubit} \bm{$\ket{\psi}$} & \textbf{Measurement bit $(B)$} &\textbf{Results ($\%$)}  \\ \hline
        \multirow{2}{*}{$\ket{0}$}      &0      &91.1    \\ \cline{2-3}
                                        &1      &8.9 \\ \hline
        \multirow{2}{*}{$\ket{1}$}      &0      &5.6    \\ \cline{2-3}
                                        &1      &94.4 \\ \hline
        \end{tabular}
        }
    \label{tab:realpercentage}
    \end{table}

As observed in the real circuit computation results (see Table~\ref{tab:realpercentage}), there are expected errors in the measurement results due to the noise in the qubits and gates in a real quantum circuit. 
While the real performance is slightly worse than the simulator with zero errors, it shows that practical medium-free secure quantum communication is possible under the QRE interpretation.

\section{Discussion} 
\label{sec:discuss} 
In this section, we discuss the implications and potential communication network structure suggested by our interpretation of remote entanglement. We also identify some of its applications in medium-free quantum communication and the associated obstacles.

\vspace{3mm} 
\noindent {\bf Existing Quantum Communication.}\ \ 
Existing modern quantum communication networks depend on multiple intermediary relay nodes and trusted parties~\cite{Liao_2017,PhysRevLett.120.030501}. 
As such, the suggested direction for a future practical quantum wide-area network would be compatible with diverse topological structures that connect distributed users in a large-scale area~\cite{chen2021integrated}. However, the widely distributed relay structure poses a few significant limitations. Most importantly, this network infrastructure is prohibitively costly to implement; the inherent capacity of the physical channels limits network transmission rates. 
Our suggested remote quantum entanglement notion and its following proposed communication protocol show that building an advanced quantum communication network without a complex topological structure could be feasible, avoiding the astronomical infrastructure set-up costs and underlying channel limitations.

\vspace{3mm} 
\noindent {\bf Potential Communication Networks.}\ \ 
In our quantum remote entanglement (QRE) interpretation, the proposed medium-free communication protocol potentially reduces the number of trusted nodes in a quantum communication network. This is because every qubit, each with a distinct identity, can directly interact with other qubits through an immediate interaction, without an existing connection or medium. Through this process, different qubits can communicate directly without any intermediaries. As a result, the number of transmission links in any given established communication could be greatly reduced to only the communicating parties involved. Thereby, communication networks can be effectively consolidated to a few central nodes, eliminating the excess cost to maintain intermediary nodes and circumvent the inherent transmission limitations of communication channels. It would be interesting to investigate further the practicality and efficacy of this consolidated network infrastructure.

\vspace{3mm} 
\noindent {\bf Information Security.}\ \ 
Following a wide-area quantum network, the direct result is a large attack surface area, each added device in the system potentially adding to the size of the attack vector. This is due to the dormant vulnerabilities in realistic devices. Moreover, any communication network built on a physical infrastructure must maintain security against known~\cite{Lo_2014,xu2020secure} and potential attacks. Several such attacks have been studied in recent years, such as blinding~\cite{lydersen2010hacking}, photon-number-splitting~\cite{brassard2000limitations}, wavelength-dependent~\cite{PhysRevA.84.062308}, Trojan-horse~\cite{PhysRevA.73.022320}, and time-shift~\cite{qi2006time} attacks.

As a direct outcome of the suggested QRE interpretation to achieve medium-less point-to-point direct quantum communication, we believe it adds a layer of strong security against the attacks as mentioned above on communication systems. The lack of intermediaries to link the communication between nodes, which potentially reduces the attack vector surface to a minimum, causes a significant issue for an adversary. The adversary would have to directly attack the endpoints of the communication system, which is much easier for the communicating parties to detect and mitigate by initiating their defense protocols. As a result, we could study how such an uncompromising network structure impacts related attacks and devise specific implementations of corresponding countermeasures, thereby increasing network security.

\vspace{3mm} 
\noindent {\bf Relevant Limitations.}\ \ 
Each qubit is represented by complex numbers $\alpha$ and $\beta$ of real continuous values. These distinguishable values define its identity. 
Given our medium-free quantum communication proposal, when Alice has a prepared qubit in a quantum state $\ket{\psi}$ to communicate to Bob, it requires that she has another qubit $A$ and the identity $QuID(B)$ of Bob's qubit $B$. Using the identity $QuID(B)$, can she remotely entangle her qubit $A$ with Bob's qubit $B$. 
Once $A$ and $B$ are entangled, she can initiate the canonical teleportation protocol~\cite{PhysRevLett.70.1895} for secure quantum communication solely based on QRE, without any physical channel for the transmission of information.
To guarantee reliable connectivity, the identity of Bob's qubit must be identified accurately, which currently depends on the resolving power of the equipment (e.g., quantum microscope, photometer).
However, this reliability can be improved with better measurement of a quantum state through the advancement of quantum tomography procedures, and it is not an inherent limitation of QRE.


\section{Related Work}
\label{sec:related}
In this section, we examine some of the most relevant works in the physical layer of communication networks and related network security issues. 

\subsection{Optical Network Security}
The use of optical fiber as a physical medium is the most common in current data transmission networks. 
While there are many types of optical networks, such as local area networks, wide area networks, and networks forming the backbone of the Internet, they are all vulnerable at the physical level in terms of security. We can broadly classify the security threats at the physical layer of fiber optic networks into a few groups. These categories of threats breach the primary information security components of confidentiality, integrity, and availability. 
A few recent survey papers~\cite{5749282,Maslo2021,7537185,7608266} identified some of the most common types of optical network attacks, including jamming, physical infrastructure attacks, eavesdropping, and interceptions.

\vspace{3mm} 
\noindent{\bf Confidentiality.} 
Data transmitted through light signals in optical fibers can be easily compromised by various eavesdropping~\cite{GUAN201831,6876451,7038129} and passive analysis~\cite{10.1117/12.883550} methods. Some attacks include (1) the physical tapping~\cite{1494884} of the fiber optics and (2) signal leakage~\cite{5533431} from significant levels of residual crosstalk in optical couplers and their components.
The tapping of optical fiber is not complicated. An insider eavesdropping attack is one of the simplest types of tapping. 
An attacker can either regularly eavesdrop on adjacent communication channels through specialized equipment or tap on switch ports of Dense Wave Division Multiplexing nodes~\cite{Maslo2021}. 

Moreover, these types of attacks are covert as they passively analyze the traffic, leaving no trace of any impact. They can be easily implemented by directly placing a second fiber adjacent to the first fiber. There is no protective material, allowing the attacker to capture a small amount of the desired optical signal from where light escapes. 
Listening to leaked signals from adjacent crosstalk channels is another form of eavesdropping. 
These attacks are performed on wavelength-division-multiplexing (WDM) networks, which cannot maintain perfect channel isolation, resulting in a small amount of optical power leakage from adjacent channels (interchannel crosstalk)~\cite{5533431}. Eavesdroppers then extract weak optical channel information using optical measurement equipment from the crosstalk. 
However, in practice, the optical fibers are protected by multiple layers of protective cabling, reducing leakage and physical tapping risk. Besides, important information is usually secured by encryption, enhancing a network's confidentiality in the physical layer~\cite{6419362}.

\vspace{3mm} 
\noindent{\bf Integrity.} 
Another requirement of secure communications is to ensure the integrity of original messages. 
Any attack that changes the data content violates data integrity. In a survey that reviewed the fundamental aspects of physical layer security, Shakiba-Herfehet al.~\cite{shakibaherfeh2020physical} identified some types of such attacks. For example, in substitution attacks, the attacker changes the message content transmitted by a legitimate source. In another example, they recognized impersonation attacks, where the attacker sends a fake message. Simultaneously, the source is idle, and the receiver should be able to detect the fake and modified messages from the authentic ones~\cite{6620768}. 
In another work, Tippenhauer et al.~\cite{10.1016/j.comnet.2016.06.021} considered the physical layer message manipulation attacks, in which an attacker changes the physical-layer properties of an original wireless message. The targeted properties are message characteristics such as time-synchronization, distance measurement, time-of-arrival, and signal strength. Because such attacks do not directly change the message content, they introduced the notion of physical-layer message integrity, which describes the absence of manipulations for physical-layer message characteristics.

\vspace{3mm} 
\noindent{\bf Availability.} 
Optical networks are highly vulnerable to attacks typically aimed at disrupting the service, gaining unauthorized access to carried data~\cite{7537185}, or causing physical infrastructure damage. 
One type of attack, jamming attacks~\cite{6876451,Rejeb2006FaultAA,1561232,8535155}, can result in a denial of service. Even though these attacks are not usually designed to steal information, they can cause significant losses of network resources. Jamming attacks are usually launched by inserting a relatively high-powered signal over either a frequency within the transmission window (in-band jamming) or out of the transmission band (out-of-band jamming) of legitimate data channels. Both types of jamming are common and can disrupt communication or even steal information with the help of a crosstalk mechanism~\cite{9013238}.


\subsection{Physical Layer Security for Wireless Networks}
The increasing prevalence of wireless devices and communications poses new challenges for physical layer security~\cite{7120011,6739367,7539590,8622294}. A cardinal characteristic of the wireless medium is its broadcasting nature. This broadcasting nature of wireless communications makes it difficult to shield transmitted signals from unintended recipients~\cite{6739367}. Another distinct feature of wireless communication is the overlapping of multiple signals at the receiver, resulting in a superposition of signals. Due to these fundamental characteristics, wireless communication networks are particularly vulnerable to eavesdropping and impersonation attacks~\cite{7539590}. 
Compared to complex cryptographic measures implemented at higher layers, physical layer security is quick and straightforward to realize.

With the recent advent and evolution toward the Internet of Things (IoT), we are embracing the future of 5G wireless technologies, which is a key driver of the growing IoT networks. However, it poses a new set of challenges for physical layer security~\cite{8335290,7081071,2020arXiv200608044V,8758230}. In such dense heterogeneous networks consisting of nodes with different transmit powers, coverage areas, and radio access technologies, it is important to identify a suitable topology to accommodate them. The IoT network is a massive network of physical objects embedded with sensors, software, and connectivity technologies, allowing them to interact among the connected devices. Vega Sánchez et al.~\cite{2020arXiv200608044V} identified the physical layer security as a promising approach that can benefit traditional encryption methods, which takes advantage of the propagation medium’s features and impairments to ensure secure communication. Yang et al.~\cite{7081071} list three promising physical layer solutions to meet the 5G security requirements. One uses a multi-tier hierarchical architecture, another deploys a massive multiple-input multiple-output, while another uses a huge swath of millimeter-wave spectrum.

\section{Conclusion}
\label{sec:conclude}
%
In this work, we identify the limitations of current quantum communication systems and consider what ought to be the resolution of quantum theory in relation to these challenges. Since many have written on the subject of remote entanglement, it may be thought presumptuous to also write of it; the more so, because in our treatment of it, we depart from the views that others have taken. However, since it is our objective to sidestep these limitations, it seems appropriate that we introduce our proposed remote quantum entanglement (QRE) interpretation, allowing for a more desirable quantum communication approach.

Modern quantum communication networks, primarily built on physical infrastructure, deploy trusted relays (e.g., quantum repeaters, low-Earth-orbit satellite) connected by optical fiber cables as a transmission medium. As a result, quantum information passing through the physical medium is subject to transmission loss and physical network layer security attacks. 
By introducing our QRE interpretation, where each qubit has a distinct quantum identity QuID and can interact from any distance using this identity, we propose a medium-free secure quantum communication approach based on the canonical quantum teleportation protocol. In the remote entanglement, assuming that QRE holds and qubits can freely entangle using their identities $QuID$s, we establish secure communication (where a sender remotely entangles with the receiver and sends the information qubit $\ket{\psi}$ in a prepared state) with a teleportation protocol. Simulations and actual quantum computations on a quantum circuit support the proposed medium-free secure quantum communication approach.






\newpage
\bibliographystyle{ACM-Reference-Format}
\bibliography{bibliography}


\begin{thebibliography}{87}


\ifx \showCODEN    \undefined \def \showCODEN     #1{\unskip}     \fi
\ifx \showDOI      \undefined \def \showDOI       #1{#1}\fi
\ifx \showISBNx    \undefined \def \showISBNx     #1{\unskip}     \fi
\ifx \showISBNxiii \undefined \def \showISBNxiii  #1{\unskip}     \fi
\ifx \showISSN     \undefined \def \showISSN      #1{\unskip}     \fi
\ifx \showLCCN     \undefined \def \showLCCN      #1{\unskip}     \fi
\ifx \shownote     \undefined \def \shownote      #1{#1}          \fi
\ifx \showarticletitle \undefined \def \showarticletitle #1{#1}   \fi
\ifx \showURL      \undefined \def \showURL       {\relax}        \fi
\providecommand\bibfield[2]{#2}
\providecommand\bibinfo[2]{#2}
\providecommand\natexlab[1]{#1}
\providecommand\showeprint[2][]{arXiv:#2}

\bibitem[\protect\citeauthoryear{Anderson, Ensher, Matthews, Wieman, and
  Cornell}{Anderson et~al\mbox{.}}{1995}]%
        {Anderson198}
\bibfield{author}{\bibinfo{person}{M.~H. Anderson}, \bibinfo{person}{J.~R.
  Ensher}, \bibinfo{person}{M.~R. Matthews}, \bibinfo{person}{C.~E. Wieman},
  {and} \bibinfo{person}{E.~A. Cornell}.} \bibinfo{year}{1995}\natexlab{}.
\newblock \showarticletitle{Observation of Bose-Einstein Condensation in a
  Dilute Atomic Vapor}.
\newblock \bibinfo{journal}{\emph{Science}} (\bibinfo{year}{1995}).
\newblock


\bibitem[\protect\citeauthoryear{Asfaw, Bello, Ben-Haim, Bozzo-Rey, Bravyi,
  Bronn, Capelluto, Vazquez, Ceroni, Chen, Frisch, Gambetta, Garion, Gil,
  Gonzalez, Harkins, Imamichi, Kang, h.~Karamlou, Loredo, McKay, Mezzacapo,
  Minev, Movassagh, Nannicni, Nation, Phan, Pistoia, Rattew, Schaefer, Shabani,
  Smolin, Stenger, Temme, Tod, Wood, and Wootton.}{Asfaw et~al\mbox{.}}{2020}]%
        {Qiskit-Textbook}
\bibfield{author}{\bibinfo{person}{Abraham Asfaw}, \bibinfo{person}{Luciano
  Bello}, \bibinfo{person}{Yael Ben-Haim}, \bibinfo{person}{Mehdi Bozzo-Rey},
  \bibinfo{person}{Sergey Bravyi}, \bibinfo{person}{Nicholas Bronn},
  \bibinfo{person}{Lauren Capelluto}, \bibinfo{person}{Almudena~Carrera
  Vazquez}, \bibinfo{person}{Jack Ceroni}, \bibinfo{person}{Richard Chen},
  \bibinfo{person}{Albert Frisch}, \bibinfo{person}{Jay Gambetta},
  \bibinfo{person}{Shelly Garion}, \bibinfo{person}{Leron Gil},
  \bibinfo{person}{Salvador De La~Puente Gonzalez}, \bibinfo{person}{Francis
  Harkins}, \bibinfo{person}{Takashi Imamichi}, \bibinfo{person}{Hwajung Kang},
  \bibinfo{person}{Amir h. Karamlou}, \bibinfo{person}{Robert Loredo},
  \bibinfo{person}{David McKay}, \bibinfo{person}{Antonio Mezzacapo},
  \bibinfo{person}{Zlatko Minev}, \bibinfo{person}{Ramis Movassagh},
  \bibinfo{person}{Giacomo Nannicni}, \bibinfo{person}{Paul Nation},
  \bibinfo{person}{Anna Phan}, \bibinfo{person}{Marco Pistoia},
  \bibinfo{person}{Arthur Rattew}, \bibinfo{person}{Joachim Schaefer},
  \bibinfo{person}{Javad Shabani}, \bibinfo{person}{John Smolin},
  \bibinfo{person}{John Stenger}, \bibinfo{person}{Kristan Temme},
  \bibinfo{person}{Madeleine Tod}, \bibinfo{person}{Stephen Wood}, {and}
  \bibinfo{person}{James Wootton.}} \bibinfo{year}{2020}\natexlab{}.
\newblock \bibinfo{booktitle}{\emph{Learn Quantum Computation Using Qiskit}}.
\newblock
\urldef\tempurl%
\url{http://community.qiskit.org/textbook}
\showURL{%
\tempurl}


\bibitem[\protect\citeauthoryear{Aspect, Grangier, and Roger}{Aspect
  et~al\mbox{.}}{1982}]%
        {aspect1982experimental}
\bibfield{author}{\bibinfo{person}{Alain Aspect}, \bibinfo{person}{Philippe
  Grangier}, {and} \bibinfo{person}{G{\'e}rard Roger}.}
  \bibinfo{year}{1982}\natexlab{}.
\newblock \showarticletitle{Experimental realization of
  Einstein-Podolsky-Rosen-Bohm Gedankenexperiment: a new violation of Bell's
  inequalities}.
\newblock \bibinfo{journal}{\emph{Physical review letters}}
  (\bibinfo{year}{1982}).
\newblock


\bibitem[\protect\citeauthoryear{Axline, Burkhart, Pfaff, Zhang, Chou,
  Campagne-Ibarcq, Reinhold, Frunzio, Girvin, Jiang, et~al\mbox{.}}{Axline
  et~al\mbox{.}}{2018}]%
        {axline2018demand}
\bibfield{author}{\bibinfo{person}{Christopher~J Axline},
  \bibinfo{person}{Luke~D Burkhart}, \bibinfo{person}{Wolfgang Pfaff},
  \bibinfo{person}{Mengzhen Zhang}, \bibinfo{person}{Kevin Chou},
  \bibinfo{person}{Philippe Campagne-Ibarcq}, \bibinfo{person}{Philip
  Reinhold}, \bibinfo{person}{Luigi Frunzio}, \bibinfo{person}{SM Girvin},
  \bibinfo{person}{Liang Jiang}, {et~al\mbox{.}}}
  \bibinfo{year}{2018}\natexlab{}.
\newblock \showarticletitle{On-demand quantum state transfer and entanglement
  between remote microwave cavity memories}.
\newblock \bibinfo{journal}{\emph{Nature Physics}} (\bibinfo{year}{2018}).
\newblock


\bibitem[\protect\citeauthoryear{{Banjac}, {Orlic}, {Peric}, and
  {Milićević}}{{Banjac} et~al\mbox{.}}{2012}]%
        {6419362}
\bibfield{author}{\bibinfo{person}{Z. {Banjac}}, \bibinfo{person}{V. {Orlic}},
  \bibinfo{person}{M. {Peric}}, {and} \bibinfo{person}{S. {Milićević}}.}
  \bibinfo{year}{2012}\natexlab{}.
\newblock \showarticletitle{Securing data on fiber optic transmission lines}.
  In \bibinfo{booktitle}{\emph{2012 20th Telecommunications Forum (TELFOR)}}.
\newblock


\bibitem[\protect\citeauthoryear{Barnes, Shilton, and Robinson}{Barnes
  et~al\mbox{.}}{2000}]%
        {PhysRevB.62.8410}
\bibfield{author}{\bibinfo{person}{C.~H.~W. Barnes}, \bibinfo{person}{J.~M.
  Shilton}, {and} \bibinfo{person}{A.~M. Robinson}.}
  \bibinfo{year}{2000}\natexlab{}.
\newblock \showarticletitle{Quantum computation using electrons trapped by
  surface acoustic waves}.
\newblock \bibinfo{journal}{\emph{Phys. Rev. B}} (\bibinfo{year}{2000}).
\newblock


\bibitem[\protect\citeauthoryear{Barnum, Caves, Fuchs, Jozsa, and
  Schumacher}{Barnum et~al\mbox{.}}{1996}]%
        {Barnum_1996}
\bibfield{author}{\bibinfo{person}{Howard Barnum}, \bibinfo{person}{Carlton~M.
  Caves}, \bibinfo{person}{Christopher~A. Fuchs}, \bibinfo{person}{Richard
  Jozsa}, {and} \bibinfo{person}{Benjamin Schumacher}.}
  \bibinfo{year}{1996}\natexlab{}.
\newblock \showarticletitle{Noncommuting Mixed States Cannot Be Broadcast}.
\newblock \bibinfo{journal}{\emph{Physical Review Letters}}
  (\bibinfo{year}{1996}).
\newblock


\bibitem[\protect\citeauthoryear{Bennett, Brassard, Cr\'epeau, Jozsa, Peres,
  and Wootters}{Bennett et~al\mbox{.}}{1993}]%
        {PhysRevLett.70.1895}
\bibfield{author}{\bibinfo{person}{Charles~H. Bennett}, \bibinfo{person}{Gilles
  Brassard}, \bibinfo{person}{Claude Cr\'epeau}, \bibinfo{person}{Richard
  Jozsa}, \bibinfo{person}{Asher Peres}, {and} \bibinfo{person}{William~K.
  Wootters}.} \bibinfo{year}{1993}\natexlab{}.
\newblock \showarticletitle{Teleporting an unknown quantum state via dual
  classical and Einstein-Podolsky-Rosen channels}.
\newblock \bibinfo{journal}{\emph{Phys. Rev. Lett.}} (\bibinfo{year}{1993}).
\newblock


\bibitem[\protect\citeauthoryear{{Bensalem}, {Singh}, and {Jukan}}{{Bensalem}
  et~al\mbox{.}}{2019}]%
        {9013238}
\bibfield{author}{\bibinfo{person}{M. {Bensalem}}, \bibinfo{person}{S.~K.
  {Singh}}, {and} \bibinfo{person}{A. {Jukan}}.}
  \bibinfo{year}{2019}\natexlab{}.
\newblock \showarticletitle{On Detecting and Preventing Jamming Attacks with
  Machine Learning in Optical Networks}. In \bibinfo{booktitle}{\emph{2019 IEEE
  Global Communications Conference (GLOBECOM)}}.
\newblock


\bibitem[\protect\citeauthoryear{Bienfait, Satzinger, Zhong, Chang, Chou,
  Conner, Dumur, Grebel, Peairs, Povey, et~al\mbox{.}}{Bienfait
  et~al\mbox{.}}{2019}]%
        {bienfait2019phonon}
\bibfield{author}{\bibinfo{person}{Audrey Bienfait}, \bibinfo{person}{Kevin~J
  Satzinger}, \bibinfo{person}{YP Zhong}, \bibinfo{person}{H-S Chang},
  \bibinfo{person}{M-H Chou}, \bibinfo{person}{Chris~R Conner},
  \bibinfo{person}{{\'E} Dumur}, \bibinfo{person}{Joel Grebel},
  \bibinfo{person}{Gregory~A Peairs}, \bibinfo{person}{Rhys~G Povey},
  {et~al\mbox{.}}} \bibinfo{year}{2019}\natexlab{}.
\newblock \showarticletitle{Phonon-mediated quantum state transfer and remote
  qubit entanglement}.
\newblock \bibinfo{journal}{\emph{Science}} (\bibinfo{year}{2019}).
\newblock


\bibitem[\protect\citeauthoryear{Bloch}{Bloch}{1946}]%
        {PhysRev.70.460}
\bibfield{author}{\bibinfo{person}{F. Bloch}.} \bibinfo{year}{1946}\natexlab{}.
\newblock \showarticletitle{Nuclear Induction}.
\newblock \bibinfo{journal}{\emph{Phys. Rev.}} (\bibinfo{year}{1946}).
\newblock


\bibitem[\protect\citeauthoryear{Bose}{Bose}{1924}]%
        {bose1924plancks}
\bibfield{author}{\bibinfo{person}{Satyendra~Nath Bose}.}
  \bibinfo{year}{1924}\natexlab{}.
\newblock \showarticletitle{Plancks gesetz und lichtquantenhypothese}.
\newblock \bibinfo{journal}{\emph{Zeitschrift für Physik}}
  (\bibinfo{year}{1924}).
\newblock


\bibitem[\protect\citeauthoryear{Brassard, L{\"u}tkenhaus, Mor, and
  Sanders}{Brassard et~al\mbox{.}}{2000}]%
        {brassard2000limitations}
\bibfield{author}{\bibinfo{person}{Gilles Brassard}, \bibinfo{person}{Norbert
  L{\"u}tkenhaus}, \bibinfo{person}{Tal Mor}, {and} \bibinfo{person}{Barry~C
  Sanders}.} \bibinfo{year}{2000}\natexlab{}.
\newblock \showarticletitle{Limitations on practical quantum cryptography}.
\newblock \bibinfo{journal}{\emph{Physical review letters}}
  (\bibinfo{year}{2000}).
\newblock


\bibitem[\protect\citeauthoryear{Campagne-Ibarcq, Zalys-Geller, Narla, Shankar,
  Reinhold, Burkhart, Axline, Pfaff, Frunzio, Schoelkopf,
  et~al\mbox{.}}{Campagne-Ibarcq et~al\mbox{.}}{2018}]%
        {campagne2018deterministic}
\bibfield{author}{\bibinfo{person}{P Campagne-Ibarcq}, \bibinfo{person}{E
  Zalys-Geller}, \bibinfo{person}{A Narla}, \bibinfo{person}{S Shankar},
  \bibinfo{person}{P Reinhold}, \bibinfo{person}{L Burkhart},
  \bibinfo{person}{C Axline}, \bibinfo{person}{W Pfaff}, \bibinfo{person}{L
  Frunzio}, \bibinfo{person}{RJ Schoelkopf}, {et~al\mbox{.}}}
  \bibinfo{year}{2018}\natexlab{}.
\newblock \showarticletitle{Deterministic remote entanglement of
  superconducting circuits through microwave two-photon transitions}.
\newblock \bibinfo{journal}{\emph{Physical review letters}}
  (\bibinfo{year}{2018}).
\newblock


\bibitem[\protect\citeauthoryear{Cavaliere, Prati, Poti, Muhammad, and
  Catuogno}{Cavaliere et~al\mbox{.}}{2020}]%
        {quantum2010007}
\bibfield{author}{\bibinfo{person}{Fabio Cavaliere}, \bibinfo{person}{Enrico
  Prati}, \bibinfo{person}{Luca Poti}, \bibinfo{person}{Imran Muhammad}, {and}
  \bibinfo{person}{Tommaso Catuogno}.} \bibinfo{year}{2020}\natexlab{}.
\newblock \showarticletitle{Secure Quantum Communication Technologies and
  Systems: From Labs to Markets}.
\newblock \bibinfo{journal}{\emph{Quantum Reports}} (\bibinfo{year}{2020}).
\newblock


\bibitem[\protect\citeauthoryear{Chang, Zhong, Bienfait, Chou, Conner, Dumur,
  Grebel, Peairs, Povey, Satzinger, et~al\mbox{.}}{Chang et~al\mbox{.}}{2020}]%
        {chang2020remote}
\bibfield{author}{\bibinfo{person}{H-S Chang}, \bibinfo{person}{YP Zhong},
  \bibinfo{person}{Audrey Bienfait}, \bibinfo{person}{M-H Chou},
  \bibinfo{person}{Christopher~R Conner}, \bibinfo{person}{{\'E}tienne Dumur},
  \bibinfo{person}{Joel Grebel}, \bibinfo{person}{Gregory~A Peairs},
  \bibinfo{person}{Rhys~G Povey}, \bibinfo{person}{Kevin~J Satzinger},
  {et~al\mbox{.}}} \bibinfo{year}{2020}\natexlab{}.
\newblock \showarticletitle{Remote entanglement via adiabatic passage using a
  tunably dissipative quantum communication system}.
\newblock \bibinfo{journal}{\emph{Physical Review Letters}}
  (\bibinfo{year}{2020}).
\newblock


\bibitem[\protect\citeauthoryear{Chen, Wang, Liang, Liu, Liu, Jiang, Wang, Wan,
  Cai, Ju, Chen, Wang, Gao, Chen, Peng, Chen, and Pan}{Chen
  et~al\mbox{.}}{2010}]%
        {Chen:2010}
\bibfield{author}{\bibinfo{person}{Teng-Yun Chen}, \bibinfo{person}{Jian Wang},
  \bibinfo{person}{Hao Liang}, \bibinfo{person}{Wei-Yue Liu},
  \bibinfo{person}{Yang Liu}, \bibinfo{person}{Xiao Jiang},
  \bibinfo{person}{Yuan Wang}, \bibinfo{person}{Xu Wan},
  \bibinfo{person}{Wen-Qi Cai}, \bibinfo{person}{Lei Ju},
  \bibinfo{person}{Luo-Kan Chen}, \bibinfo{person}{Liu-Jun Wang},
  \bibinfo{person}{Yuan Gao}, \bibinfo{person}{Kai Chen},
  \bibinfo{person}{Cheng-Zhi Peng}, \bibinfo{person}{Zeng-Bing Chen}, {and}
  \bibinfo{person}{Jian-Wei Pan}.} \bibinfo{year}{2010}\natexlab{}.
\newblock \showarticletitle{Metropolitan all-pass and inter-city quantum
  communication network}.
\newblock \bibinfo{journal}{\emph{Opt. Express}} (\bibinfo{year}{2010}).
\newblock


\bibitem[\protect\citeauthoryear{Chen, Zhang, Chen, Cai, Liao, Zhang, Chen,
  Yin, Ren, Chen, et~al\mbox{.}}{Chen et~al\mbox{.}}{2021}]%
        {chen2021integrated}
\bibfield{author}{\bibinfo{person}{Yu-Ao Chen}, \bibinfo{person}{Qiang Zhang},
  \bibinfo{person}{Teng-Yun Chen}, \bibinfo{person}{Wen-Qi Cai},
  \bibinfo{person}{Sheng-Kai Liao}, \bibinfo{person}{Jun Zhang},
  \bibinfo{person}{Kai Chen}, \bibinfo{person}{Juan Yin},
  \bibinfo{person}{Ji-Gang Ren}, \bibinfo{person}{Zhu Chen}, {et~al\mbox{.}}}
  \bibinfo{year}{2021}\natexlab{}.
\newblock \showarticletitle{An integrated space-to-ground quantum communication
  network over 4,600 kilometres}.
\newblock \bibinfo{journal}{\emph{Nature}} (\bibinfo{year}{2021}).
\newblock


\bibitem[\protect\citeauthoryear{Dahlberg, Skrzypczyk, Coopmans, Wubben,
  Rozpundefineddek, Pompili, Stolk, Pawe\l{}czak, Knegjens, de~Oliveira~Filho,
  Hanson, and Wehner}{Dahlberg et~al\mbox{.}}{2019}]%
        {10.1145/3341302.3342070}
\bibfield{author}{\bibinfo{person}{Axel Dahlberg}, \bibinfo{person}{Matthew
  Skrzypczyk}, \bibinfo{person}{Tim Coopmans}, \bibinfo{person}{Leon Wubben},
  \bibinfo{person}{Filip Rozpundefineddek}, \bibinfo{person}{Matteo Pompili},
  \bibinfo{person}{Arian Stolk}, \bibinfo{person}{Przemys\l{}aw Pawe\l{}czak},
  \bibinfo{person}{Robert Knegjens}, \bibinfo{person}{Julio de Oliveira~Filho},
  \bibinfo{person}{Ronald Hanson}, {and} \bibinfo{person}{Stephanie Wehner}.}
  \bibinfo{year}{2019}\natexlab{}.
\newblock \showarticletitle{A Link Layer Protocol for Quantum Networks}. In
  \bibinfo{booktitle}{\emph{Proceedings of the ACM Special Interest Group on
  Data Communication}}.
\newblock


\bibitem[\protect\citeauthoryear{De~Broglie}{De~Broglie}{1924}]%
        {de1924recherches}
\bibfield{author}{\bibinfo{person}{Louis De~Broglie}.}
  \bibinfo{year}{1924}\natexlab{}.
\newblock \emph{\bibinfo{title}{Recherches sur la th{\'e}orie des quanta}}.
\newblock \bibinfo{thesistype}{Ph.\,D. Dissertation}.
  \bibinfo{school}{Migration-universit{\'e} en cours d'affectation}.
\newblock


\bibitem[\protect\citeauthoryear{Diamanti, Lo, Qi, and Yuan}{Diamanti
  et~al\mbox{.}}{2016}]%
        {Diamanti_2016}
\bibfield{author}{\bibinfo{person}{Eleni Diamanti}, \bibinfo{person}{Hoi-Kwong
  Lo}, \bibinfo{person}{Bing Qi}, {and} \bibinfo{person}{Zhiliang Yuan}.}
  \bibinfo{year}{2016}\natexlab{}.
\newblock \showarticletitle{Practical challenges in quantum key distribution}.
\newblock \bibinfo{journal}{\emph{npj Quantum Information}}
  (\bibinfo{year}{2016}).
\newblock


\bibitem[\protect\citeauthoryear{Dickel, Wesdorp, Langford, Peiter,
  Sagastizabal, Bruno, Criger, Motzoi, and DiCarlo}{Dickel
  et~al\mbox{.}}{2018}]%
        {dickel2018chip}
\bibfield{author}{\bibinfo{person}{Christian Dickel}, \bibinfo{person}{JJ
  Wesdorp}, \bibinfo{person}{NK Langford}, \bibinfo{person}{S Peiter},
  \bibinfo{person}{Ramiro Sagastizabal}, \bibinfo{person}{Alessandro Bruno},
  \bibinfo{person}{Ben Criger}, \bibinfo{person}{F Motzoi}, {and}
  \bibinfo{person}{L DiCarlo}.} \bibinfo{year}{2018}\natexlab{}.
\newblock \showarticletitle{Chip-to-chip entanglement of transmon qubits using
  engineered measurement fields}.
\newblock \bibinfo{journal}{\emph{Physical Review B}} (\bibinfo{year}{2018}).
\newblock


\bibitem[\protect\citeauthoryear{Dieks}{Dieks}{1982}]%
        {DIEKS1982271}
\bibfield{author}{\bibinfo{person}{D. Dieks}.} \bibinfo{year}{1982}\natexlab{}.
\newblock \showarticletitle{Communication by EPR devices}.
\newblock \bibinfo{journal}{\emph{Physics Letters A}} (\bibinfo{year}{1982}).
\newblock


\bibitem[\protect\citeauthoryear{Duan, Lukin, Cirac, and Zoller}{Duan
  et~al\mbox{.}}{2001}]%
        {duan2001long}
\bibfield{author}{\bibinfo{person}{L-M Duan}, \bibinfo{person}{Mikhail~D
  Lukin}, \bibinfo{person}{J~Ignacio Cirac}, {and} \bibinfo{person}{Peter
  Zoller}.} \bibinfo{year}{2001}\natexlab{}.
\newblock \showarticletitle{Long-distance quantum communication with atomic
  ensembles and linear optics}.
\newblock \bibinfo{journal}{\emph{Nature}} (\bibinfo{year}{2001}).
\newblock


\bibitem[\protect\citeauthoryear{Dušek, Lütkenhaus, and Hendrych}{Dušek
  et~al\mbox{.}}{2006}]%
        {Du_ek_2006}
\bibfield{author}{\bibinfo{person}{Miloslav Dušek}, \bibinfo{person}{Norbert
  Lütkenhaus}, {and} \bibinfo{person}{Martin Hendrych}.}
  \bibinfo{year}{2006}\natexlab{}.
\newblock \showarticletitle{Quantum cryptography}.
\newblock \bibinfo{journal}{\emph{Progress in Optics}} (\bibinfo{year}{2006}).
\newblock


\bibitem[\protect\citeauthoryear{Einstein, Podolsky, and Rosen}{Einstein
  et~al\mbox{.}}{1935}]%
        {einstein1935can}
\bibfield{author}{\bibinfo{person}{Albert Einstein}, \bibinfo{person}{Boris
  Podolsky}, {and} \bibinfo{person}{Nathan Rosen}.}
  \bibinfo{year}{1935}\natexlab{}.
\newblock \showarticletitle{Can quantum-mechanical description of physical
  reality be considered complete?}
\newblock \bibinfo{journal}{\emph{Physical review}} (\bibinfo{year}{1935}).
\newblock


\bibitem[\protect\citeauthoryear{Elliott}{Elliott}{2002}]%
        {elliott2002building}
\bibfield{author}{\bibinfo{person}{Chip Elliott}.}
  \bibinfo{year}{2002}\natexlab{}.
\newblock \showarticletitle{Building the quantum network}.
\newblock \bibinfo{journal}{\emph{New Journal of Physics}}
  (\bibinfo{year}{2002}).
\newblock


\bibitem[\protect\citeauthoryear{{Fan}, {Yang}, {Duong}, {Elkashlan}, and
  {Karagiannidis}}{{Fan} et~al\mbox{.}}{2016}]%
        {7407417}
\bibfield{author}{\bibinfo{person}{L. {Fan}}, \bibinfo{person}{N. {Yang}},
  \bibinfo{person}{T.~Q. {Duong}}, \bibinfo{person}{M. {Elkashlan}}, {and}
  \bibinfo{person}{G.~K. {Karagiannidis}}.} \bibinfo{year}{2016}\natexlab{}.
\newblock \showarticletitle{Exploiting Direct Links for Physical Layer Security
  in Multiuser Multirelay Networks}.
\newblock \bibinfo{journal}{\emph{IEEE Transactions on Wireless
  Communications}} (\bibinfo{year}{2016}).
\newblock


\bibitem[\protect\citeauthoryear{{Fok}, {Wang}, {Deng}, and {Prucnal}}{{Fok}
  et~al\mbox{.}}{2011}]%
        {5749282}
\bibfield{author}{\bibinfo{person}{M.~P. {Fok}}, \bibinfo{person}{Z. {Wang}},
  \bibinfo{person}{Y. {Deng}}, {and} \bibinfo{person}{P.~R. {Prucnal}}.}
  \bibinfo{year}{2011}\natexlab{}.
\newblock \showarticletitle{Optical Layer Security in Fiber-Optic Networks}.
\newblock \bibinfo{journal}{\emph{IEEE Transactions on Information Forensics
  and Security}} (\bibinfo{year}{2011}).
\newblock


\bibitem[\protect\citeauthoryear{{Furdek}, {Skorin-Kapov}, {Bosiljevac}, and
  {Šipuš}}{{Furdek} et~al\mbox{.}}{2010}]%
        {5533431}
\bibfield{author}{\bibinfo{person}{M. {Furdek}}, \bibinfo{person}{N.
  {Skorin-Kapov}}, \bibinfo{person}{M. {Bosiljevac}}, {and} \bibinfo{person}{Z.
  {Šipuš}}.} \bibinfo{year}{2010}\natexlab{}.
\newblock \showarticletitle{Analysis of crosstalk in optical couplers and
  associated vulnerabilities}. In \bibinfo{booktitle}{\emph{The 33rd
  International Convention MIPRO}}.
\newblock


\bibitem[\protect\citeauthoryear{{Furdek}, {Skorin-Kapov}, {Zsigmond}, and
  {Wosinska}}{{Furdek} et~al\mbox{.}}{2014}]%
        {6876451}
\bibfield{author}{\bibinfo{person}{M. {Furdek}}, \bibinfo{person}{N.
  {Skorin-Kapov}}, \bibinfo{person}{S. {Zsigmond}}, {and} \bibinfo{person}{L.
  {Wosinska}}.} \bibinfo{year}{2014}\natexlab{}.
\newblock \showarticletitle{Vulnerabilities and security issues in optical
  networks}. In \bibinfo{booktitle}{\emph{2014 16th International Conference on
  Transparent Optical Networks (ICTON)}}.
\newblock


\bibitem[\protect\citeauthoryear{{Furdek}, {Wosinska}, {Goścień},
  {Manousakis}, {Aibin}, {Walkowiak}, {Ristov}, {Gushev}, and {Marzo}}{{Furdek}
  et~al\mbox{.}}{2016}]%
        {7608266}
\bibfield{author}{\bibinfo{person}{M. {Furdek}}, \bibinfo{person}{L.
  {Wosinska}}, \bibinfo{person}{R. {Goścień}}, \bibinfo{person}{K.
  {Manousakis}}, \bibinfo{person}{M. {Aibin}}, \bibinfo{person}{K.
  {Walkowiak}}, \bibinfo{person}{S. {Ristov}}, \bibinfo{person}{M. {Gushev}},
  {and} \bibinfo{person}{J.~L. {Marzo}}.} \bibinfo{year}{2016}\natexlab{}.
\newblock \showarticletitle{An overview of security challenges in communication
  networks}. In \bibinfo{booktitle}{\emph{2016 8th International Workshop on
  Resilient Networks Design and Modeling (RNDM)}}.
\newblock


\bibitem[\protect\citeauthoryear{Gisin, Fasel, Kraus, Zbinden, and
  Ribordy}{Gisin et~al\mbox{.}}{2006}]%
        {PhysRevA.73.022320}
\bibfield{author}{\bibinfo{person}{N. Gisin}, \bibinfo{person}{S. Fasel},
  \bibinfo{person}{B. Kraus}, \bibinfo{person}{H. Zbinden}, {and}
  \bibinfo{person}{G. Ribordy}.} \bibinfo{year}{2006}\natexlab{}.
\newblock \showarticletitle{Trojan-horse attacks on quantum-key-distribution
  systems}.
\newblock \bibinfo{journal}{\emph{Phys. Rev. A}} (\bibinfo{year}{2006}).
\newblock


\bibitem[\protect\citeauthoryear{Gisin, Ribordy, Tittel, and Zbinden}{Gisin
  et~al\mbox{.}}{2002}]%
        {RevModPhys.74.145}
\bibfield{author}{\bibinfo{person}{Nicolas Gisin}, \bibinfo{person}{Gr\'egoire
  Ribordy}, \bibinfo{person}{Wolfgang Tittel}, {and} \bibinfo{person}{Hugo
  Zbinden}.} \bibinfo{year}{2002}\natexlab{}.
\newblock \showarticletitle{Quantum cryptography}.
\newblock \bibinfo{journal}{\emph{Rev. Mod. Phys.}} (\bibinfo{year}{2002}).
\newblock


\bibitem[\protect\citeauthoryear{Gisin and Thew}{Gisin and Thew}{2007}]%
        {gisin2007quantum}
\bibfield{author}{\bibinfo{person}{Nicolas Gisin} {and} \bibinfo{person}{Rob
  Thew}.} \bibinfo{year}{2007}\natexlab{}.
\newblock \showarticletitle{Quantum communication}.
\newblock \bibinfo{journal}{\emph{Nature photonics}} (\bibinfo{year}{2007}).
\newblock


\bibitem[\protect\citeauthoryear{Guan, Cho, and Winzer}{Guan
  et~al\mbox{.}}{2018}]%
        {GUAN201831}
\bibfield{author}{\bibinfo{person}{Kyle Guan}, \bibinfo{person}{Junho Cho},
  {and} \bibinfo{person}{Peter~J. Winzer}.} \bibinfo{year}{2018}\natexlab{}.
\newblock \showarticletitle{Physical layer security in fiber-optic MIMO-SDM
  systems: An overview}.
\newblock \bibinfo{journal}{\emph{Optics Communications}}
  (\bibinfo{year}{2018}).
\newblock


\bibitem[\protect\citeauthoryear{Hassan and Hijazi}{Hassan and Hijazi}{2017}]%
        {HASSAN2017133}
\bibfield{author}{\bibinfo{person}{Nihad~Ahmad Hassan} {and}
  \bibinfo{person}{Rami Hijazi}.} \bibinfo{year}{2017}\natexlab{}.
\newblock \showarticletitle{Chapter 5 - Data Hiding Using Encryption
  Techniques}.
\newblock In \bibinfo{booktitle}{\emph{Data Hiding Techniques in Windows OS}}.
\newblock


\bibitem[\protect\citeauthoryear{Hensen, Bernien, Dr{\'e}au, Reiserer, Kalb,
  Blok, Ruitenberg, Vermeulen, Schouten, Abell{\'a}n, et~al\mbox{.}}{Hensen
  et~al\mbox{.}}{2015}]%
        {hensen2015loophole}
\bibfield{author}{\bibinfo{person}{Bas Hensen}, \bibinfo{person}{Hannes
  Bernien}, \bibinfo{person}{Ana{\"\i}s~E Dr{\'e}au}, \bibinfo{person}{Andreas
  Reiserer}, \bibinfo{person}{Norbert Kalb}, \bibinfo{person}{Machiel~S Blok},
  \bibinfo{person}{Just Ruitenberg}, \bibinfo{person}{Raymond~FL Vermeulen},
  \bibinfo{person}{Raymond~N Schouten}, \bibinfo{person}{Carlos Abell{\'a}n},
  {et~al\mbox{.}}} \bibinfo{year}{2015}\natexlab{}.
\newblock \showarticletitle{Loophole-free Bell inequality violation using
  electron spins separated by 1.3 kilometres}.
\newblock \bibinfo{journal}{\emph{Nature}} (\bibinfo{year}{2015}).
\newblock


\bibitem[\protect\citeauthoryear{Huelga, Vaccaro, Chefles, and Plenio}{Huelga
  et~al\mbox{.}}{2001}]%
        {huelga2001quantum}
\bibfield{author}{\bibinfo{person}{Susana~F Huelga}, \bibinfo{person}{Joan~A
  Vaccaro}, \bibinfo{person}{Anthony Chefles}, {and} \bibinfo{person}{Martin~B
  Plenio}.} \bibinfo{year}{2001}\natexlab{}.
\newblock \showarticletitle{Quantum remote control: teleportation of unitary
  operations}.
\newblock \bibinfo{journal}{\emph{Physical Review A}} (\bibinfo{year}{2001}).
\newblock


\bibitem[\protect\citeauthoryear{Humphreys, Kalb, Morits, Schouten, Vermeulen,
  Twitchen, Markham, and Hanson}{Humphreys et~al\mbox{.}}{2018}]%
        {humphreys2018deterministic}
\bibfield{author}{\bibinfo{person}{Peter~C Humphreys}, \bibinfo{person}{Norbert
  Kalb}, \bibinfo{person}{Jaco~PJ Morits}, \bibinfo{person}{Raymond~N
  Schouten}, \bibinfo{person}{Raymond~FL Vermeulen}, \bibinfo{person}{Daniel~J
  Twitchen}, \bibinfo{person}{Matthew Markham}, {and} \bibinfo{person}{Ronald
  Hanson}.} \bibinfo{year}{2018}\natexlab{}.
\newblock \showarticletitle{Deterministic delivery of remote entanglement on a
  quantum network}.
\newblock \bibinfo{journal}{\emph{Nature}} (\bibinfo{year}{2018}).
\newblock


\bibitem[\protect\citeauthoryear{Inagaki, Matsuda, Tadanaga, Asobe, and
  Takesue}{Inagaki et~al\mbox{.}}{2013}]%
        {Inagaki_2013}
\bibfield{author}{\bibinfo{person}{Takahiro Inagaki}, \bibinfo{person}{Nobuyuki
  Matsuda}, \bibinfo{person}{Osamu Tadanaga}, \bibinfo{person}{Masaki Asobe},
  {and} \bibinfo{person}{Hiroki Takesue}.} \bibinfo{year}{2013}\natexlab{}.
\newblock \showarticletitle{Entanglement distribution over 300 km of fiber}.
\newblock \bibinfo{journal}{\emph{Optics Express}} (\bibinfo{year}{2013}).
\newblock


\bibitem[\protect\citeauthoryear{Julsgaard, Kozhekin, and Polzik}{Julsgaard
  et~al\mbox{.}}{2001}]%
        {julsgaard2001experimental}
\bibfield{author}{\bibinfo{person}{Brian Julsgaard}, \bibinfo{person}{Alexander
  Kozhekin}, {and} \bibinfo{person}{Eugene~S Polzik}.}
  \bibinfo{year}{2001}\natexlab{}.
\newblock \showarticletitle{Experimental long-lived entanglement of two
  macroscopic objects}.
\newblock \bibinfo{journal}{\emph{Nature}} (\bibinfo{year}{2001}).
\newblock


\bibitem[\protect\citeauthoryear{Kessler}{Kessler}{2003}]%
        {kessler2003overview}
\bibfield{author}{\bibinfo{person}{Gary~C Kessler}.}
  \bibinfo{year}{2003}\natexlab{}.
\newblock \showarticletitle{An overview of cryptography}.
\newblock \bibinfo{journal}{\emph{[Online]. Available:
  https://www.garykessler.net/library/crypto.html}} (\bibinfo{year}{2003}).
\newblock


\bibitem[\protect\citeauthoryear{Kurose and Ross}{Kurose and Ross}{2012}]%
        {kurosecomputer}
\bibfield{author}{\bibinfo{person}{James~F. Kurose} {and}
  \bibinfo{person}{Keith~W. Ross}.} \bibinfo{year}{2012}\natexlab{}.
\newblock \bibinfo{booktitle}{\emph{Computer Networking: A Top-Down Approach
  (6th Edition)}}.
\newblock \bibinfo{publisher}{Pearson}.
\newblock


\bibitem[\protect\citeauthoryear{Kurpiers, Magnard, Walter, Royer, Pechal,
  Heinsoo, Salath{\'e}, Akin, Storz, Besse, et~al\mbox{.}}{Kurpiers
  et~al\mbox{.}}{2018}]%
        {kurpiers2018deterministic}
\bibfield{author}{\bibinfo{person}{Philipp Kurpiers}, \bibinfo{person}{Paul
  Magnard}, \bibinfo{person}{Theo Walter}, \bibinfo{person}{Baptiste Royer},
  \bibinfo{person}{Marek Pechal}, \bibinfo{person}{Johannes Heinsoo},
  \bibinfo{person}{Yves Salath{\'e}}, \bibinfo{person}{Abdulkadir Akin},
  \bibinfo{person}{Simon Storz}, \bibinfo{person}{J-C Besse}, {et~al\mbox{.}}}
  \bibinfo{year}{2018}\natexlab{}.
\newblock \showarticletitle{Deterministic quantum state transfer and remote
  entanglement using microwave photons}.
\newblock \bibinfo{journal}{\emph{Nature}} (\bibinfo{year}{2018}).
\newblock


\bibitem[\protect\citeauthoryear{Leung, Lu, Chakram, Naik, Earnest, Ma, Jacobs,
  Cleland, and Schuster}{Leung et~al\mbox{.}}{2019}]%
        {leung2019deterministic}
\bibfield{author}{\bibinfo{person}{N Leung}, \bibinfo{person}{Y Lu},
  \bibinfo{person}{S Chakram}, \bibinfo{person}{RK Naik}, \bibinfo{person}{N
  Earnest}, \bibinfo{person}{R Ma}, \bibinfo{person}{K Jacobs},
  \bibinfo{person}{AN Cleland}, {and} \bibinfo{person}{DI Schuster}.}
  \bibinfo{year}{2019}\natexlab{}.
\newblock \showarticletitle{Deterministic bidirectional communication and
  remote entanglement generation between superconducting qubits}.
\newblock \bibinfo{journal}{\emph{npj Quantum Information}}
  (\bibinfo{year}{2019}).
\newblock


\bibitem[\protect\citeauthoryear{Li, Wang, Huang, Chen, Yin, Li, Zhou, Liu,
  Zhang, Guo, Bao, and Han}{Li et~al\mbox{.}}{2011}]%
        {PhysRevA.84.062308}
\bibfield{author}{\bibinfo{person}{Hong-Wei Li}, \bibinfo{person}{Shuang Wang},
  \bibinfo{person}{Jing-Zheng Huang}, \bibinfo{person}{Wei Chen},
  \bibinfo{person}{Zhen-Qiang Yin}, \bibinfo{person}{Fang-Yi Li},
  \bibinfo{person}{Zheng Zhou}, \bibinfo{person}{Dong Liu},
  \bibinfo{person}{Yang Zhang}, \bibinfo{person}{Guang-Can Guo},
  \bibinfo{person}{Wan-Su Bao}, {and} \bibinfo{person}{Zheng-Fu Han}.}
  \bibinfo{year}{2011}\natexlab{}.
\newblock \showarticletitle{Attacking a practical quantum-key-distribution
  system with wavelength-dependent beam-splitter and multiwavelength sources}.
\newblock \bibinfo{journal}{\emph{Phys. Rev. A}} (\bibinfo{year}{2011}).
\newblock


\bibitem[\protect\citeauthoryear{Liao, Cai, Handsteiner, Liu, Yin, Zhang,
  Rauch, Fink, Ren, Liu, Li, Shen, Cao, Li, Wang, Huang, Deng, Xi, Ma, Hu, Li,
  Liu, Koidl, Wang, Chen, Wang, Steindorfer, Kirchner, Lu, Shu, Ursin, Scheidl,
  Peng, Wang, Zeilinger, and Pan}{Liao et~al\mbox{.}}{2018}]%
        {PhysRevLett.120.030501}
\bibfield{author}{\bibinfo{person}{Sheng-Kai Liao}, \bibinfo{person}{Wen-Qi
  Cai}, \bibinfo{person}{Johannes Handsteiner}, \bibinfo{person}{Bo Liu},
  \bibinfo{person}{Juan Yin}, \bibinfo{person}{Liang Zhang},
  \bibinfo{person}{Dominik Rauch}, \bibinfo{person}{Matthias Fink},
  \bibinfo{person}{Ji-Gang Ren}, \bibinfo{person}{Wei-Yue Liu},
  \bibinfo{person}{Yang Li}, \bibinfo{person}{Qi Shen}, \bibinfo{person}{Yuan
  Cao}, \bibinfo{person}{Feng-Zhi Li}, \bibinfo{person}{Jian-Feng Wang},
  \bibinfo{person}{Yong-Mei Huang}, \bibinfo{person}{Lei Deng},
  \bibinfo{person}{Tao Xi}, \bibinfo{person}{Lu Ma}, \bibinfo{person}{Tai Hu},
  \bibinfo{person}{Li Li}, \bibinfo{person}{Nai-Le Liu}, \bibinfo{person}{Franz
  Koidl}, \bibinfo{person}{Peiyuan Wang}, \bibinfo{person}{Yu-Ao Chen},
  \bibinfo{person}{Xiang-Bin Wang}, \bibinfo{person}{Michael Steindorfer},
  \bibinfo{person}{Georg Kirchner}, \bibinfo{person}{Chao-Yang Lu},
  \bibinfo{person}{Rong Shu}, \bibinfo{person}{Rupert Ursin},
  \bibinfo{person}{Thomas Scheidl}, \bibinfo{person}{Cheng-Zhi Peng},
  \bibinfo{person}{Jian-Yu Wang}, \bibinfo{person}{Anton Zeilinger}, {and}
  \bibinfo{person}{Jian-Wei Pan}.} \bibinfo{year}{2018}\natexlab{}.
\newblock \showarticletitle{Satellite-Relayed Intercontinental Quantum
  Network}.
\newblock \bibinfo{journal}{\emph{Phys. Rev. Lett.}} (\bibinfo{year}{2018}).
\newblock


\bibitem[\protect\citeauthoryear{Liao, Cai, Liu, Zhang, Li, Ren, Yin, Shen,
  Cao, Li, and et~al.}{Liao et~al\mbox{.}}{2017}]%
        {Liao_2017}
\bibfield{author}{\bibinfo{person}{Sheng-Kai Liao}, \bibinfo{person}{Wen-Qi
  Cai}, \bibinfo{person}{Wei-Yue Liu}, \bibinfo{person}{Liang Zhang},
  \bibinfo{person}{Yang Li}, \bibinfo{person}{Ji-Gang Ren},
  \bibinfo{person}{Juan Yin}, \bibinfo{person}{Qi Shen}, \bibinfo{person}{Yuan
  Cao}, \bibinfo{person}{Zheng-Ping Li}, {and} \bibinfo{person}{et al.}}
  \bibinfo{year}{2017}\natexlab{}.
\newblock \showarticletitle{Satellite-to-ground quantum key distribution}.
\newblock \bibinfo{journal}{\emph{Nature}} (\bibinfo{year}{2017}).
\newblock


\bibitem[\protect\citeauthoryear{{Liu}, {Chen}, and {Wang}}{{Liu}
  et~al\mbox{.}}{2017}]%
        {7539590}
\bibfield{author}{\bibinfo{person}{Y. {Liu}}, \bibinfo{person}{H. {Chen}},
  {and} \bibinfo{person}{L. {Wang}}.} \bibinfo{year}{2017}\natexlab{}.
\newblock \showarticletitle{Physical Layer Security for Next Generation
  Wireless Networks: Theories, Technologies, and Challenges}.
\newblock \bibinfo{journal}{\emph{IEEE Communications Surveys Tutorials}}
  (\bibinfo{year}{2017}).
\newblock


\bibitem[\protect\citeauthoryear{Lo, Curty, and Tamaki}{Lo
  et~al\mbox{.}}{2014}]%
        {Lo_2014}
\bibfield{author}{\bibinfo{person}{Hoi-Kwong Lo}, \bibinfo{person}{Marcos
  Curty}, {and} \bibinfo{person}{Kiyoshi Tamaki}.}
  \bibinfo{year}{2014}\natexlab{}.
\newblock \showarticletitle{Secure quantum key distribution}.
\newblock \bibinfo{journal}{\emph{Nature Photonics}} (\bibinfo{year}{2014}).
\newblock


\bibitem[\protect\citeauthoryear{Lo and Zhao}{Lo and Zhao}{2008}]%
        {lo2008quantum}
\bibfield{author}{\bibinfo{person}{Hoi-Kwong Lo} {and} \bibinfo{person}{Yi
  Zhao}.} \bibinfo{year}{2008}\natexlab{}.
\newblock \showarticletitle{Quantum Cryptography}.
\newblock \bibinfo{journal}{\emph{arXiv}}  \bibinfo{volume}{0803.2507}
  (\bibinfo{year}{2008}).
\newblock


\bibitem[\protect\citeauthoryear{{Lopez-Martinez}, {Gomez}, and
  {Garrido-Balsells}}{{Lopez-Martinez} et~al\mbox{.}}{2015}]%
        {7038129}
\bibfield{author}{\bibinfo{person}{F.~J. {Lopez-Martinez}}, \bibinfo{person}{G.
  {Gomez}}, {and} \bibinfo{person}{J.~M. {Garrido-Balsells}}.}
  \bibinfo{year}{2015}\natexlab{}.
\newblock \showarticletitle{Physical-Layer Security in Free-Space Optical
  Communications}.
\newblock \bibinfo{journal}{\emph{IEEE Photonics Journal}}
  (\bibinfo{year}{2015}).
\newblock


\bibitem[\protect\citeauthoryear{Lv, Zhao, and Zhou}{Lv et~al\mbox{.}}{2018}]%
        {lv2018joint}
\bibfield{author}{\bibinfo{person}{Shu-Xin Lv}, \bibinfo{person}{Zheng-Wei
  Zhao}, {and} \bibinfo{person}{Ping Zhou}.} \bibinfo{year}{2018}\natexlab{}.
\newblock \showarticletitle{Joint remote control of an arbitrary single-qubit
  state by using a multiparticle entangled state as the quantum channel}.
\newblock \bibinfo{journal}{\emph{Quantum Information Processing}}
  (\bibinfo{year}{2018}).
\newblock


\bibitem[\protect\citeauthoryear{Lvovsky and Raymer}{Lvovsky and
  Raymer}{2009}]%
        {lvovsky2009continuous}
\bibfield{author}{\bibinfo{person}{Alexander~I Lvovsky} {and}
  \bibinfo{person}{Michael~G Raymer}.} \bibinfo{year}{2009}\natexlab{}.
\newblock \showarticletitle{Continuous-variable optical quantum-state
  tomography}.
\newblock \bibinfo{journal}{\emph{Reviews of modern physics}}
  (\bibinfo{year}{2009}).
\newblock


\bibitem[\protect\citeauthoryear{Lydersen, Wiechers, Wittmann, Elser, Skaar,
  and Makarov}{Lydersen et~al\mbox{.}}{2010}]%
        {lydersen2010hacking}
\bibfield{author}{\bibinfo{person}{Lars Lydersen}, \bibinfo{person}{Carlos
  Wiechers}, \bibinfo{person}{Christoffer Wittmann}, \bibinfo{person}{Dominique
  Elser}, \bibinfo{person}{Johannes Skaar}, {and} \bibinfo{person}{Vadim
  Makarov}.} \bibinfo{year}{2010}\natexlab{}.
\newblock \showarticletitle{Hacking commercial quantum cryptography systems by
  tailored bright illumination}.
\newblock \bibinfo{journal}{\emph{Nature photonics}} (\bibinfo{year}{2010}).
\newblock


\bibitem[\protect\citeauthoryear{Maslo, Sarajli{\'{c}}, Hod{\v{z}}i{\'{c}}, and
  Muj{\v{c}}i{\'{c}}}{Maslo et~al\mbox{.}}{2021}]%
        {Maslo2021}
\bibfield{author}{\bibinfo{person}{Anis Maslo}, \bibinfo{person}{Nermin
  Sarajli{\'{c}}}, \bibinfo{person}{Mujo Hod{\v{z}}i{\'{c}}}, {and}
  \bibinfo{person}{Aljo Muj{\v{c}}i{\'{c}}}.} \bibinfo{year}{2021}\natexlab{}.
\newblock \bibinfo{booktitle}{\emph{Optical Network Security Attacks by Tapping
  and Encrypting Optical Signals}}.
\newblock \bibinfo{publisher}{Springer International Publishing}.
\newblock


\bibitem[\protect\citeauthoryear{{Mukherjee}, {Fakoorian}, {Huang}, and
  {Swindlehurst}}{{Mukherjee} et~al\mbox{.}}{2014}]%
        {6739367}
\bibfield{author}{\bibinfo{person}{A. {Mukherjee}}, \bibinfo{person}{S.~A.~A.
  {Fakoorian}}, \bibinfo{person}{J. {Huang}}, {and} \bibinfo{person}{A.~L.
  {Swindlehurst}}.} \bibinfo{year}{2014}\natexlab{}.
\newblock \showarticletitle{Principles of Physical Layer Security in Multiuser
  Wireless Networks: A Survey}.
\newblock \bibinfo{journal}{\emph{IEEE Communications Surveys Tutorials}}
  (\bibinfo{year}{2014}).
\newblock


\bibitem[\protect\citeauthoryear{Narla, Shankar, Hatridge, Leghtas, Sliwa,
  Zalys-Geller, Mundhada, Pfaff, Frunzio, Schoelkopf, et~al\mbox{.}}{Narla
  et~al\mbox{.}}{2016}]%
        {narla2016robust}
\bibfield{author}{\bibinfo{person}{Anirudh Narla}, \bibinfo{person}{Shyam
  Shankar}, \bibinfo{person}{Michael Hatridge}, \bibinfo{person}{Zaki Leghtas},
  \bibinfo{person}{Katrina~M Sliwa}, \bibinfo{person}{Evan Zalys-Geller},
  \bibinfo{person}{Shantanu~O Mundhada}, \bibinfo{person}{Wolfgang Pfaff},
  \bibinfo{person}{Luigi Frunzio}, \bibinfo{person}{Robert~J Schoelkopf},
  {et~al\mbox{.}}} \bibinfo{year}{2016}\natexlab{}.
\newblock \showarticletitle{Robust concurrent remote entanglement between two
  superconducting qubits}.
\newblock \bibinfo{journal}{\emph{Physical Review X}} (\bibinfo{year}{2016}).
\newblock


\bibitem[\protect\citeauthoryear{{Natalino}, {Schiano}, {Di Giglio},
  {Wosinska}, and {Furdek}}{{Natalino} et~al\mbox{.}}{2018}]%
        {8535155}
\bibfield{author}{\bibinfo{person}{C. {Natalino}}, \bibinfo{person}{M.
  {Schiano}}, \bibinfo{person}{A. {Di Giglio}}, \bibinfo{person}{L.
  {Wosinska}}, {and} \bibinfo{person}{M. {Furdek}}.}
  \bibinfo{year}{2018}\natexlab{}.
\newblock \showarticletitle{Field Demonstration of Machine-Learning-Aided
  Detection and Identification of Jamming Attacks in Optical Networks}. In
  \bibinfo{booktitle}{\emph{2018 European Conference on Optical Communication
  (ECOC)}}.
\newblock


\bibitem[\protect\citeauthoryear{{Obeed}, {Salhab}, {Alouini}, and
  {Zummo}}{{Obeed} et~al\mbox{.}}{2018}]%
        {8622294}
\bibfield{author}{\bibinfo{person}{M. {Obeed}}, \bibinfo{person}{A.~M.
  {Salhab}}, \bibinfo{person}{M. {Alouini}}, {and} \bibinfo{person}{S.~A.
  {Zummo}}.} \bibinfo{year}{2018}\natexlab{}.
\newblock \showarticletitle{Survey on Physical Layer Security in Optical
  Wireless Communication Systems}. In \bibinfo{booktitle}{\emph{2018 Seventh
  International Conference on Communications and Networking (ComNet)}}.
\newblock


\bibitem[\protect\citeauthoryear{{Perlaza}, {Chorti}, {Poor}, and
  {Han}}{{Perlaza} et~al\mbox{.}}{2013}]%
        {6620768}
\bibfield{author}{\bibinfo{person}{S.~M. {Perlaza}}, \bibinfo{person}{A.
  {Chorti}}, \bibinfo{person}{H.~V. {Poor}}, {and} \bibinfo{person}{Z. {Han}}.}
  \bibinfo{year}{2013}\natexlab{}.
\newblock \showarticletitle{On the impact of network-state knowledge on the
  Feasibility of secrecy}. In \bibinfo{booktitle}{\emph{2013 IEEE International
  Symposium on Information Theory}}.
\newblock


\bibitem[\protect\citeauthoryear{Qi, Fung, Lo, and Ma}{Qi
  et~al\mbox{.}}{2006}]%
        {qi2006time}
\bibfield{author}{\bibinfo{person}{Bing Qi}, \bibinfo{person}{Chi-Hang Fung},
  \bibinfo{person}{Hoi-Kwong Lo}, {and} \bibinfo{person}{Xiongfeng Ma}.}
  \bibinfo{year}{2006}\natexlab{}.
\newblock \showarticletitle{Time-shift Attack in Practical Quantum
  Cryptosystems}.
\newblock \bibinfo{journal}{\emph{Quantum Information \& Computation}}
  (\bibinfo{year}{2006}).
\newblock


\bibitem[\protect\citeauthoryear{Rashed and Tabbour}{Rashed and
  Tabbour}{2017}]%
        {rashed2017suitable}
\bibfield{author}{\bibinfo{person}{Ahmed Nabih~Zaki Rashed} {and}
  \bibinfo{person}{Mohammed Salah~F Tabbour}.} \bibinfo{year}{2017}\natexlab{}.
\newblock \showarticletitle{Suitable optical fiber communication channel for
  optical nonlinearity signal processing in high optical data rate systems}.
\newblock \bibinfo{journal}{\emph{Wireless Personal Communications}}
  (\bibinfo{year}{2017}).
\newblock


\bibitem[\protect\citeauthoryear{Rejeb, Leeson, and Green}{Rejeb
  et~al\mbox{.}}{2006}]%
        {Rejeb2006FaultAA}
\bibfield{author}{\bibinfo{person}{R. Rejeb}, \bibinfo{person}{M. Leeson},
  {and} \bibinfo{person}{R. Green}.} \bibinfo{year}{2006}\natexlab{}.
\newblock \showarticletitle{Fault and attack management in all-optical
  networks}.
\newblock \bibinfo{journal}{\emph{IEEE Communications Magazine}}
  (\bibinfo{year}{2006}).
\newblock


\bibitem[\protect\citeauthoryear{Roch, Schwartz, Motzoi, Macklin, Vijay,
  Eddins, Korotkov, Whaley, Sarovar, and Siddiqi}{Roch et~al\mbox{.}}{2014}]%
        {PhysRevLett.112.170501}
\bibfield{author}{\bibinfo{person}{N. Roch}, \bibinfo{person}{M.~E. Schwartz},
  \bibinfo{person}{F. Motzoi}, \bibinfo{person}{C. Macklin},
  \bibinfo{person}{R. Vijay}, \bibinfo{person}{A.~W. Eddins},
  \bibinfo{person}{A.~N. Korotkov}, \bibinfo{person}{K.~B. Whaley},
  \bibinfo{person}{M. Sarovar}, {and} \bibinfo{person}{I. Siddiqi}.}
  \bibinfo{year}{2014}\natexlab{}.
\newblock \showarticletitle{Observation of Measurement-Induced Entanglement and
  Quantum Trajectories of Remote Superconducting Qubits}.
\newblock \bibinfo{journal}{\emph{Phys. Rev. Lett.}} (\bibinfo{year}{2014}).
\newblock


\bibitem[\protect\citeauthoryear{Scarani, Bechmann-Pasquinucci, Cerf,
  Du\ifmmode~\check{s}\else \v{s}\fi{}ek, L\"utkenhaus, and Peev}{Scarani
  et~al\mbox{.}}{2009}]%
        {RevModPhys.81.1301}
\bibfield{author}{\bibinfo{person}{Valerio Scarani}, \bibinfo{person}{Helle
  Bechmann-Pasquinucci}, \bibinfo{person}{Nicolas~J. Cerf},
  \bibinfo{person}{Miloslav Du\ifmmode~\check{s}\else \v{s}\fi{}ek},
  \bibinfo{person}{Norbert L\"utkenhaus}, {and} \bibinfo{person}{Momtchil
  Peev}.} \bibinfo{year}{2009}\natexlab{}.
\newblock \showarticletitle{The security of practical quantum key
  distribution}.
\newblock \bibinfo{journal}{\emph{Rev. Mod. Phys.}} (\bibinfo{year}{2009}).
\newblock


\bibitem[\protect\citeauthoryear{Sch{\"u}tz}{Sch{\"u}tz}{2017}]%
        {schutz2017universal}
\bibfield{author}{\bibinfo{person}{Martin~JA Sch{\"u}tz}.}
  \bibinfo{year}{2017}\natexlab{}.
\newblock \showarticletitle{Universal quantum transducers based on surface
  acoustic waves}.
\newblock In \bibinfo{booktitle}{\emph{Quantum Dots for Quantum Information
  Processing: Controlling and Exploiting the Quantum Dot Environment}}.
\newblock


\bibitem[\protect\citeauthoryear{Shakiba-Herfeh, Chorti, and
  Poor}{Shakiba-Herfeh et~al\mbox{.}}{2020}]%
        {shakibaherfeh2020physical}
\bibfield{author}{\bibinfo{person}{Mahdi Shakiba-Herfeh},
  \bibinfo{person}{Arsenia Chorti}, {and} \bibinfo{person}{H.~Vince Poor}.}
  \bibinfo{year}{2020}\natexlab{}.
\newblock \showarticletitle{Physical Layer Security: Authentication, Integrity
  and Confidentiality}.
\newblock \bibinfo{journal}{\emph{arXiv}}  \bibinfo{volume}{2001.07153}
  (\bibinfo{year}{2020}).
\newblock


\bibitem[\protect\citeauthoryear{{Shaneman} and {Gray}}{{Shaneman} and
  {Gray}}{2004}]%
        {1494884}
\bibfield{author}{\bibinfo{person}{K. {Shaneman}} {and} \bibinfo{person}{S.
  {Gray}}.} \bibinfo{year}{2004}\natexlab{}.
\newblock \showarticletitle{Optical network security: technical analysis of
  fiber tapping mechanisms and methods for detection prevention}. In
  \bibinfo{booktitle}{\emph{IEEE MILCOM 2004. Military Communications
  Conference, 2004.}}
\newblock


\bibitem[\protect\citeauthoryear{Simon}{Simon}{2002}]%
        {PhysRevA.66.052323}
\bibfield{author}{\bibinfo{person}{Christoph Simon}.}
  \bibinfo{year}{2002}\natexlab{}.
\newblock \showarticletitle{Natural entanglement in Bose-Einstein condensates}.
\newblock \bibinfo{journal}{\emph{Phys. Rev. A}} (\bibinfo{year}{2002}).
\newblock


\bibitem[\protect\citeauthoryear{{Skorin-Kapov}, {Furdek}, {Zsigmond}, and
  {Wosinska}}{{Skorin-Kapov} et~al\mbox{.}}{2016}]%
        {7537185}
\bibfield{author}{\bibinfo{person}{N. {Skorin-Kapov}}, \bibinfo{person}{M.
  {Furdek}}, \bibinfo{person}{S. {Zsigmond}}, {and} \bibinfo{person}{L.
  {Wosinska}}.} \bibinfo{year}{2016}\natexlab{}.
\newblock \showarticletitle{Physical-layer security in evolving optical
  networks}.
\newblock \bibinfo{journal}{\emph{IEEE Communications Magazine}}
  (\bibinfo{year}{2016}).
\newblock


\bibitem[\protect\citeauthoryear{{Tao Wu} and {Somani}}{{Tao Wu} and
  {Somani}}{2005}]%
        {1561232}
\bibfield{author}{\bibinfo{person}{{Tao Wu}} {and} \bibinfo{person}{A.~K.
  {Somani}}.} \bibinfo{year}{2005}\natexlab{}.
\newblock \showarticletitle{Cross-talk attack monitoring and localization in
  all-optical networks}.
\newblock \bibinfo{journal}{\emph{IEEE/ACM Transactions on Networking}}
  (\bibinfo{year}{2005}).
\newblock


\bibitem[\protect\citeauthoryear{Tippenhauer, Rasmussen, and
  Capkun}{Tippenhauer et~al\mbox{.}}{2016}]%
        {10.1016/j.comnet.2016.06.021}
\bibfield{author}{\bibinfo{person}{Nils~Ole Tippenhauer},
  \bibinfo{person}{Kasper~Bonne Rasmussen}, {and} \bibinfo{person}{Srdjan
  Capkun}.} \bibinfo{year}{2016}\natexlab{}.
\newblock \showarticletitle{Physical-Layer Integrity for Wireless Messages}.
\newblock \bibinfo{journal}{\emph{Comput. Netw.}} (\bibinfo{year}{2016}).
\newblock


\bibitem[\protect\citeauthoryear{Tittel, Brendel, Gisin, Herzog, Zbinden, and
  Gisin}{Tittel et~al\mbox{.}}{1998}]%
        {tittel1998experimental}
\bibfield{author}{\bibinfo{person}{Wolfgang Tittel},
  \bibinfo{person}{J{\"u}rgen Brendel}, \bibinfo{person}{Bernard Gisin},
  \bibinfo{person}{Thomas Herzog}, \bibinfo{person}{Hugo Zbinden}, {and}
  \bibinfo{person}{Nicolas Gisin}.} \bibinfo{year}{1998}\natexlab{}.
\newblock \showarticletitle{Experimental demonstration of quantum correlations
  over more than 10 km}.
\newblock \bibinfo{journal}{\emph{Physical Review A}} (\bibinfo{year}{1998}).
\newblock


\bibitem[\protect\citeauthoryear{{Trappe}}{{Trappe}}{2015}]%
        {7120011}
\bibfield{author}{\bibinfo{person}{W. {Trappe}}.}
  \bibinfo{year}{2015}\natexlab{}.
\newblock \showarticletitle{The challenges facing physical layer security}.
\newblock \bibinfo{journal}{\emph{IEEE Communications Magazine}}
  (\bibinfo{year}{2015}).
\newblock


\bibitem[\protect\citeauthoryear{Vallone, Bacco, Dequal, Gaiarin, Luceri,
  Bianco, and Villoresi}{Vallone et~al\mbox{.}}{2015}]%
        {PhysRevLett.115.040502}
\bibfield{author}{\bibinfo{person}{Giuseppe Vallone}, \bibinfo{person}{Davide
  Bacco}, \bibinfo{person}{Daniele Dequal}, \bibinfo{person}{Simone Gaiarin},
  \bibinfo{person}{Vincenza Luceri}, \bibinfo{person}{Giuseppe Bianco}, {and}
  \bibinfo{person}{Paolo Villoresi}.} \bibinfo{year}{2015}\natexlab{}.
\newblock \showarticletitle{Experimental Satellite Quantum Communications}.
\newblock \bibinfo{journal}{\emph{Phys. Rev. Lett.}} (\bibinfo{year}{2015}).
\newblock


\bibitem[\protect\citeauthoryear{{Vega S{\'a}nchez}, {Urquiza-Aguiar}, {Paredes
  Paredes}, and {Moya Osorio}}{{Vega S{\'a}nchez} et~al\mbox{.}}{2020}]%
        {2020arXiv200608044V}
\bibfield{author}{\bibinfo{person}{Jos{\'e}~David {Vega S{\'a}nchez}},
  \bibinfo{person}{Luis {Urquiza-Aguiar}}, \bibinfo{person}{Martha~Cecilia
  {Paredes Paredes}}, {and} \bibinfo{person}{Diana~Pamela {Moya Osorio}}.}
  \bibinfo{year}{2020}\natexlab{}.
\newblock \showarticletitle{{Survey on Physical Layer Security for 5G Wireless
  Networks}}.
\newblock \bibinfo{journal}{\emph{Ann. Telecommun}} (\bibinfo{year}{2020}).
\newblock


\bibitem[\protect\citeauthoryear{Vishnu, Joy, Behera, and Panigrahi}{Vishnu
  et~al\mbox{.}}{2018}]%
        {vishnu2018experimental}
\bibfield{author}{\bibinfo{person}{PK Vishnu}, \bibinfo{person}{Dintomon Joy},
  \bibinfo{person}{Bikash~K Behera}, {and} \bibinfo{person}{Prasanta~K
  Panigrahi}.} \bibinfo{year}{2018}\natexlab{}.
\newblock \showarticletitle{Experimental demonstration of non-local
  controlled-unitary quantum gates using a five-qubit quantum computer}.
\newblock \bibinfo{journal}{\emph{Quantum Information Processing}}
  (\bibinfo{year}{2018}).
\newblock


\bibitem[\protect\citeauthoryear{{Wang}, {Wang}, {Alipour-Fanid}, {Jiao}, and
  {Zeng}}{{Wang} et~al\mbox{.}}{2019}]%
        {8758230}
\bibfield{author}{\bibinfo{person}{N. {Wang}}, \bibinfo{person}{P. {Wang}},
  \bibinfo{person}{A. {Alipour-Fanid}}, \bibinfo{person}{L. {Jiao}}, {and}
  \bibinfo{person}{K. {Zeng}}.} \bibinfo{year}{2019}\natexlab{}.
\newblock \showarticletitle{Physical-Layer Security of 5G Wireless Networks for
  IoT: Challenges and Opportunities}.
\newblock \bibinfo{journal}{\emph{IEEE Internet of Things Journal}}
  (\bibinfo{year}{2019}).
\newblock


\bibitem[\protect\citeauthoryear{White and Pilbeam}{White and Pilbeam}{2011}]%
        {10.1117/12.883550}
\bibfield{author}{\bibinfo{person}{Joshua~S. White} {and}
  \bibinfo{person}{Adam~W. Pilbeam}.} \bibinfo{year}{2011}\natexlab{}.
\newblock \showarticletitle{{An analysis of coupling attacks in high-speed
  fiber optic networks}}. In \bibinfo{booktitle}{\emph{Enabling Photonics
  Technologies for Defense, Security, and Aerospace Applications VII}}.
\newblock


\bibitem[\protect\citeauthoryear{Wootters and Zurek}{Wootters and
  Zurek}{1982}]%
        {Wootters1982Single}
\bibfield{author}{\bibinfo{person}{W.~K. Wootters} {and} \bibinfo{person}{W.~H.
  Zurek}.} \bibinfo{year}{1982}\natexlab{}.
\newblock \showarticletitle{A single quantum cannot be cloned}.
\newblock \bibinfo{journal}{\emph{Nature}} (\bibinfo{year}{1982}).
\newblock


\bibitem[\protect\citeauthoryear{{Wu}, {Khisti}, {Xiao}, {Caire}, {Wong}, and
  {Gao}}{{Wu} et~al\mbox{.}}{2018}]%
        {8335290}
\bibfield{author}{\bibinfo{person}{Y. {Wu}}, \bibinfo{person}{A. {Khisti}},
  \bibinfo{person}{C. {Xiao}}, \bibinfo{person}{G. {Caire}},
  \bibinfo{person}{K. {Wong}}, {and} \bibinfo{person}{X. {Gao}}.}
  \bibinfo{year}{2018}\natexlab{}.
\newblock \showarticletitle{A Survey of Physical Layer Security Techniques for
  5G Wireless Networks and Challenges Ahead}.
\newblock \bibinfo{journal}{\emph{IEEE Journal on Selected Areas in
  Communications}} (\bibinfo{year}{2018}).
\newblock


\bibitem[\protect\citeauthoryear{Xu, Ma, Zhang, Lo, and Pan}{Xu
  et~al\mbox{.}}{2020}]%
        {xu2020secure}
\bibfield{author}{\bibinfo{person}{Feihu Xu}, \bibinfo{person}{Xiongfeng Ma},
  \bibinfo{person}{Qiang Zhang}, \bibinfo{person}{Hoi-Kwong Lo}, {and}
  \bibinfo{person}{Jian-Wei Pan}.} \bibinfo{year}{2020}\natexlab{}.
\newblock \showarticletitle{Secure quantum key distribution with realistic
  devices}.
\newblock \bibinfo{journal}{\emph{Reviews of Modern Physics}}
  (\bibinfo{year}{2020}).
\newblock


\bibitem[\protect\citeauthoryear{{Yang}, {Wang}, {Geraci}, {Elkashlan}, {Yuan},
  and {Di Renzo}}{{Yang} et~al\mbox{.}}{2015}]%
        {7081071}
\bibfield{author}{\bibinfo{person}{N. {Yang}}, \bibinfo{person}{L. {Wang}},
  \bibinfo{person}{G. {Geraci}}, \bibinfo{person}{M. {Elkashlan}},
  \bibinfo{person}{J. {Yuan}}, {and} \bibinfo{person}{M. {Di Renzo}}.}
  \bibinfo{year}{2015}\natexlab{}.
\newblock \showarticletitle{Safeguarding 5G wireless communication networks
  using physical layer security}.
\newblock \bibinfo{journal}{\emph{IEEE Communications Magazine}}
  (\bibinfo{year}{2015}).
\newblock


\bibitem[\protect\citeauthoryear{Ying-Qiao, Xing-Ri, and Shou}{Ying-Qiao
  et~al\mbox{.}}{2005}]%
        {Ying_Qiao_2005}
\bibfield{author}{\bibinfo{person}{Zhang Ying-Qiao}, \bibinfo{person}{Jin
  Xing-Ri}, {and} \bibinfo{person}{Zhang Shou}.}
  \bibinfo{year}{2005}\natexlab{}.
\newblock \showarticletitle{Probabilistic remote preparation of a two-atom
  entangled state}.
\newblock \bibinfo{journal}{\emph{Chinese Physics}} (\bibinfo{year}{2005}).
\newblock


\bibitem[\protect\citeauthoryear{Zhong, Chang, Satzinger, Chou, Bienfait,
  Conner, Dumur, Grebel, Peairs, Povey, et~al\mbox{.}}{Zhong
  et~al\mbox{.}}{2019}]%
        {zhong2019violating}
\bibfield{author}{\bibinfo{person}{YP Zhong}, \bibinfo{person}{H-S Chang},
  \bibinfo{person}{KJ Satzinger}, \bibinfo{person}{M-H Chou},
  \bibinfo{person}{Audrey Bienfait}, \bibinfo{person}{CR Conner},
  \bibinfo{person}{{\'E} Dumur}, \bibinfo{person}{Joel Grebel},
  \bibinfo{person}{GA Peairs}, \bibinfo{person}{RG Povey}, {et~al\mbox{.}}}
  \bibinfo{year}{2019}\natexlab{}.
\newblock \showarticletitle{Violating Bell’s inequality with remotely
  connected superconducting qubits}.
\newblock \bibinfo{journal}{\emph{Nature Physics}} (\bibinfo{year}{2019}).
\newblock


\end{thebibliography}

\appendix
\section*{Appendix}

\section{State Vector Simulations}
We perform additional state vector simulations, where the communication qubit $\ket{\psi}$ is prepared in various states, and report the results from running the quantum circuit for the simulations in this paper.

\subsection{State Vector \texorpdfstring{ $\ket{\psi}_{r}$}{} }

In this state vector simulation, Alice prepares the communication qubit state $\ket{\psi}_{r}$ as:
\begin{equation} \label{eq:statevec_add0}
\ket{\psi}_{r} =  
\begin{pmatrix}
\alpha    \\
\beta   \\
\end{pmatrix} = 
\begin{pmatrix}
0.24517+0.46166i    \\
0.81676+0.24426i    \\
\end{pmatrix}
\end{equation}
which in a Bloch sphere representation, $\ket{\psi}_{r}$ is visually presented as (see Figure~\ref{fig:statevec_add0}):

    \begin{figure}[htbp]
        \centering
        \includegraphics[width=.35\linewidth]{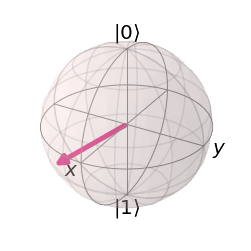}
        \caption{Bloch sphere representation of the quantum state $\ket{\psi}_{r}$.}
        \label{fig:statevec_add0}
    \end{figure}        

In the end, the quantum circuit outputs a three-qubit state vector:

        \begin{equation*}
        \substack{\ket{\psi}_{r} \\ \text{output} \\ \text{state vector}} = 
        \begin{bmatrix}
        0 \\
        0 \\
        0 \\
        0.24517 + 0.46166i \\
        0 \\
        0 \\
        0 \\
        0.81676 + 0.24426i \\        
        \end{bmatrix}
        \end{equation*}

\begin{figure}[htbp]
        \centering
        \includegraphics[width=1.0\linewidth]{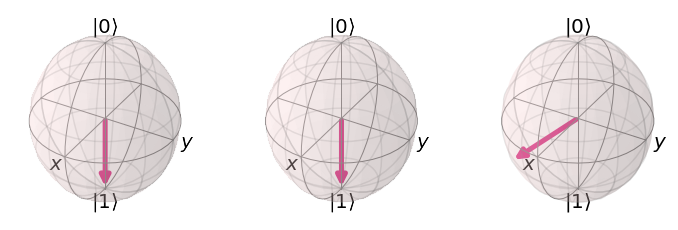}
        \caption{At the end of the circuit, both of Alice's qubits, $\ket{\psi}_{r}$ and $A$, collapse to either $\ket{0}$ or $\ket{1}$, and Bob's qubit $B$ is in the same state as the prepared communication state $\ket{\psi}_{r}$.}
        \label{fig:statevec2_add0}
    \end{figure}

Both of Alice's qubits, $\ket{\psi}_{r}$ and $A$, collapse to either $\ket{0}$ or $\ket{1}$, while Bob's qubit $B$ becomes the $\ket{\psi}_{r}$ as prepared by Alice prior to the communication; $\ket{\psi}_{r}$ has been teleported from Alice to Bob, completing the quantum communication (see Figure~\ref{fig:statevec2_add0}).

\subsection{State Vector \texorpdfstring{ $\ket{\psi}_{s}$}{} }

In this state vector simulation, Alice prepares the communication qubit state $\ket{\psi}_{s}$ as:
\begin{equation} \label{eq:statevec_add1}
\ket{\psi}_{s} =  
\begin{pmatrix}
\alpha    \\
\beta   \\
\end{pmatrix} = 
\begin{pmatrix}
0.66915 - 0.64644i    \\
0.36011 + 0.06845i    \\
\end{pmatrix}
\end{equation}
which we visually present the state $\ket{\psi}_{s}$ by plotting it in a Bloch sphere representation (see Figure~\ref{fig:statevec_add1}):

    \begin{figure}[htbp]
        \centering
        \includegraphics[width=.35\linewidth]{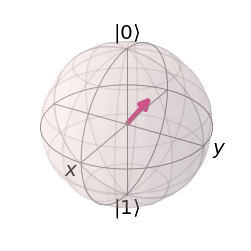}
        \caption{Bloch sphere of the quantum state $\ket{\psi}_{s}$ to be communicated.}
        \label{fig:statevec_add1}
    \end{figure}        

\pagebreak
In the end, the quantum circuit outputs a three-qubit state vector:

        \begin{equation*}
        \substack{\ket{\psi}_{s} \\ \text{output} \\ \text{state vector}} = 
        \begin{bmatrix}
        0 \\
        0 \\
        0.66915 - 0.64644i    \\
        0    \\
        0 \\
        0 \\
        0.36011 + 0.06845i    \\
        0 \\
        \end{bmatrix}
        \end{equation*}

\begin{figure}[htbp]
        \centering
        \includegraphics[width=1.0\linewidth]{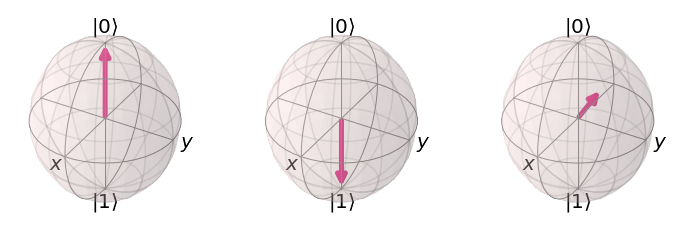}
        \caption{At the end of the circuit, both of Alice's qubits, $\ket{\psi}_{s}$ and $A$, collapse to either $\ket{0}$ or $\ket{1}$, and Bob's qubit $B$ is in the same state as the prepared communication state $\ket{\psi}_{s}$.}
        \label{fig:statevec2_add1}
    \end{figure}

Both of Alice's qubits, $\ket{\psi}_{s}$ and $A$, collapse to either $\ket{0}$ or $\ket{1}$, while Bob's qubit $B$ becomes the $\ket{\psi}_{s}$ as prepared by Alice prior to the communication. Now, $\ket{\psi}_{s}$ has been teleported from Alice to Bob, completing the quantum communication (see Figure~\ref{fig:statevec2_add1}).

\end{document}